\documentstyle[aps,pra,amsmath,amssymb,epsfig,twocolumn]{revtex}

\newcommand{\C}{{\mathbb C}} 
\newcommand{\Tr}{{\mathrm Tr}}
\newcommand{\bs}{\boldsymbol}
\newcommand{\beps}{\boldsymbol\varepsilon}
\newcommand{\bbeps}{\boldsymbol{\bar\varepsilon}}
\newcommand{\lpar}{lin\,{\parallel}\, lin} 
\newcommand{\lperp}{lin\,{\perp}\,lin}      
\newcommand{\hpar}{h\,{\parallel}\, h}     
\newcommand{\hperp}{h\,{\perp}\,h}         
\newcommand{\ket}[1]{|#1\rangle} 
\newcommand{\bra}[1]{\langle#1|}
\newcommand{\mv}[1]{\left\langle#1\right\rangle}
\newcommand{\ps}[2]{({\bf #1}\cdot{\bf #2})}
\newcommand{\Je}{J_{\mathrm e}}
\newcommand{\me}{m_{\mathrm e}}
\newcommand{\transitiontypes}{Transition types:
  ($\blacktriangledown,\triangledown$): $\Je=J+1$,
  ($\scriptstyle \blacksquare,\square$): $\Je=J$,
  ($\blacktriangle,\vartriangle$): $\Je= J-1$}

\newlength{\argwidth}
\newlength{\vertexheight}
\newcommand{\vertex}[4]{
	\settowidth{\argwidth}{$#1$}
	\hspace{\argwidth}
	\raisebox{\vertexheight}[20pt][15pt]{%
        \setlength{\unitlength}{1pt}
        \begin{picture}(25,25)(0,0)
        	\put(0,0){\line(1,0){25}}
        	\put(0,25){\line(1,0){25}}
        	\put(8.5,0){\line(0,1){25}}
        	\put(16.5,0){\line(0,1){25}}
		\put(0,25){\hspace{-\argwidth}\raisebox{-0.5ex}{$#1$}}
		\put(25,25){\raisebox{-0.5ex}{$#2$}}
		\put(25,0){\raisebox{-0.5ex}{$#3$}}
		\settowidth{\argwidth}{$#4$}	
		\put(0,0){\hspace{-\argwidth}\raisebox{-0.5ex}{$#4$}}
   	\end{picture}}
	\settowidth{\argwidth}{$#3$}
	\hspace{\argwidth}
}
\newcommand{\horizontal}[4]{
	\settowidth{\argwidth}{$#1$}
	\hspace{\argwidth}
	\raisebox{\vertexheight}[20pt][15pt]{%
        \setlength{\unitlength}{1pt}
        \begin{picture}(15,25)(0,0)
        	\put(0,0){\line(1,0){15}}
        	\put(0,25){\line(1,0){15}}
		\put(0,25){\hspace{-\argwidth}\raisebox{-0.5ex}{$#1$}}
		\put(15,25){\raisebox{-0.5ex}{$#2$}}
		\put(15,0){\raisebox{-0.5ex}{$#3$}}
		\settowidth{\argwidth}{$#4$}	
		\put(0,0){\hspace{-\argwidth}\raisebox{-0.5ex}{$#4$}}
   	\end{picture}}
	\settowidth{\argwidth}{$#3$}
	\hspace{\argwidth}
}
\newcommand{\diagonal}[4]{
	\settowidth{\argwidth}{$#1$}
	\hspace{\argwidth}
	\raisebox{\vertexheight}[20pt][15pt]{%
        \setlength{\unitlength}{1pt}
        \begin{picture}(25,25)(0,0)
		\put(0,2.5){\line(5,4){11}}
	      	\put(25,22.5){\line(-5,-4){11}}
              	\put(0,22.5){\line(5,-4){25}}
		\put(0,25){\hspace{-\argwidth}\raisebox{-0.5ex}{$#1$}}
		\put(25,25){\raisebox{-0.5ex}{$#2$}}
		\put(25,0){\raisebox{-0.5ex}{$#3$}}
		\settowidth{\argwidth}{$#4$}	
		\put(0,0){\hspace{-\argwidth}\raisebox{-0.5ex}{$#4$}}
   	\end{picture}}
	\settowidth{\argwidth}{$#3$}
	\hspace{\argwidth}
}
\newcommand{\vertical}[4]{
	\settowidth{\argwidth}{$#1$}
	\hspace{\argwidth}
	\raisebox{\vertexheight}[20pt][15pt]{%
        \setlength{\unitlength}{1pt}
        \begin{picture}(20,25)(0,0)
        	\put(0,7.5){\line(0,1){12.5}}
        	\put(20,7.5){\line(0,1){12.5}}
		\put(0,25){\hspace{-0.5\argwidth}\raisebox{-0.5ex}{$#1$}}
		\put(20,25){\hspace{-0.5\argwidth}\raisebox{-0.5ex}{$#2$}}
		\put(20,0){\hspace{-0.5\argwidth}\raisebox{-0.5ex}{$#3$}}
		\settowidth{\argwidth}{$#4$}	
		\put(0,0){\hspace{-0.5\argwidth}\raisebox{-0.5ex}{$#4$}}
   	\end{picture}}
	\settowidth{\argwidth}{$#3$}
	\hspace{\argwidth}
}

\newcommand{\clvertex}[4]{
	\settowidth{\argwidth}{$#1$}
	\hspace{\argwidth}
	\raisebox{\vertexheight}[20pt][15pt]{%
        \setlength{\unitlength}{1pt}
        \begin{picture}(25,25)(0,0)
        	\put(0,0){\line(1,0){25}}
        	\put(0,25){\line(1,0){25}}
	        \qbezier[15](12.5,0)(12.5,12.5)(12.5,25)
		\put(0,25){\hspace{-\argwidth}\raisebox{-0.5ex}{$#1$}}
		\put(25,25){\raisebox{-0.5ex}{$#2$}}
		\put(25,0){\raisebox{-0.5ex}{$#3$}}
		\settowidth{\argwidth}{$#4$}	
		\put(0,0){\hspace{-\argwidth}\raisebox{-0.5ex}{$#4$}}
   	\end{picture}}
	\settowidth{\argwidth}{$#3$}
	\hspace{\argwidth}}
\newcommand{\doubleladder}[4]{
	\vertex{#1}{\Delta}{\Delta}{#4}
	\hspace{-0.5\argwidth}
	\vertex{}{#2}{#3}{}}
\newcommand{\cldoubleladder}[4]{
	\clvertex{#1}{\Delta}{\Delta}{#4}
	\hspace{-0.5\argwidth}
	\clvertex{}{#2}{#3}{}}
\newcommand{\doublecrossed}[4]{
	\settowidth{\argwidth}{$#1$}
	\hspace{\argwidth}
	\raisebox{\vertexheight}[20pt][15pt]{%
        \setlength{\unitlength}{1pt}
        \begin{picture}(50,25)(0,0)
		\put(0,0){\line(1,0){21}}
      		\put(0,25){\line(1,0){21}}
      		\put(29,0){\line(1,0){21}}
      		\put(29,25){\line(1,0){21}}
      		\put(6,0){\line(6,5){14}}
      		\put(14,0){\line(6,5){10}}
      		\put(36,25){\line(-6,-5){10}}
      		\put(44,25){\line(-6,-5){14}}
      		\put(6,25){\line(6,-5){30}}
      		\put(14,25){\line(6,-5){30}}
		\put(0,25){\hspace{-\argwidth}\raisebox{-0.5ex}{$#1$}}
		\put(50,25){\raisebox{-0.5ex}{$#2$}}
		\put(50,0){\raisebox{-0.5ex}{$#3$}}
		\settowidth{\argwidth}{$#4$}	
		\put(0,0){\hspace{-\argwidth}\raisebox{-0.5ex}{$#4$}}
		\settowidth{\argwidth}{$\Delta$}
		\put(25,0){\hspace{-0.5\argwidth}\raisebox{-0.5ex}{$\Delta$}}
		\put(25,25){\hspace{-0.5\argwidth}\raisebox{-0.5ex}{$\Delta$}}
    	\end{picture}}}
\newcommand{\cldoublecrossed}[4]{
	\settowidth{\argwidth}{$#1$}
	\hspace{\argwidth}
	\raisebox{\vertexheight}[20pt][15pt]{%
        \setlength{\unitlength}{1pt}
        \begin{picture}(50,25)(0,0)
		\put(0,0){\line(1,0){21}}
      		\put(0,25){\line(1,0){21}}
      		\put(29,0){\line(1,0){21}}
      		\put(29,25){\line(1,0){21}}
 	     	\qbezier[20](10,0)(25,12.5)(40,25)
      		\qbezier[20](10,25)(25,12.5)(40,0)
 		\put(0,25){\hspace{-\argwidth}\raisebox{-0.5ex}{$#1$}}
		\put(50,25){\raisebox{-0.5ex}{$#2$}}
		\put(50,0){\raisebox{-0.5ex}{$#3$}}
		\settowidth{\argwidth}{$#4$}	
		\put(0,0){\hspace{-\argwidth}\raisebox{-0.5ex}{$#4$}}
		\settowidth{\argwidth}{$\Delta$}
		\put(25,0){\hspace{-0.5\argwidth}\raisebox{-0.5ex}{$\Delta$}}
		\put(25,25){\hspace{-0.5\argwidth}\raisebox{-0.5ex}{$\Delta$}}
    	\end{picture}}}
\newcommand{\twistedvertex}[4]{
	\settowidth{\argwidth}{$#1$}
	\hspace{\argwidth}
	\raisebox{\vertexheight}[20pt][15pt]{%
        \setlength{\unitlength}{1pt}
        \begin{picture}(25,25)(0,0)
        	\put(0,0){\line(1,0){25}}
        	\put(0,25){\line(1,0){25}}
	        \qbezier(8.5,0)(8.5,6.25)(12.5,12.5)
		\qbezier(12.5,12.5)(16.5,18.75)(16.5,25)
        	\qbezier(8.5,25)(8.5,18.75)(11.86,13.5)
		\qbezier(13.14,11.5)(16.5,6.25)(16.5,0)
 		\put(0,25){\hspace{-\argwidth}\raisebox{-0.5ex}{$#1$}}
		\put(25,25){\raisebox{-0.5ex}{$#2$}}
		\put(25,0){\raisebox{-0.5ex}{$#3$}}
		\settowidth{\argwidth}{$#4$}	
		\put(0,0){\hspace{-\argwidth}\raisebox{-0.5ex}{$#4$}}
   	\end{picture}}
	\settowidth{\argwidth}{$#3$}
	\hspace{\argwidth}
}
\newcommand{\twisteddoublecrossed}[4]{
	\twistedvertex{#1}{\Delta}{\Delta}{#4}
	\hspace{-0.5\argwidth}
	\twistedvertex{}{#2}{#3}{}}

\begin{document}

\author{Cord A. M\"{u}ller$^{1,3}$, Thibaut Jonckheere$^{2}$,
Christian Miniatura$^{1}$ and Dominique Delande$^{2,3}$}
\address{ $^1$ Laboratoire Ondes et D\'esordre, FRE 2302 du CNRS,
  1361 route des Lucioles,
  F-06560\ Valbonne, France\\
  $^2$ Laboratoire Kastler Brossel,
  Universit\'e Pierre
  et Marie Curie, \\
  Tour 12, Etage 1, 4 Place Jussieu,
  F-75252 Paris Cedex 05, France\\
  $^3$ Max-Planck-Institut f\"ur Physik komplexer Systeme, 
  N\"othnitzer Str. 38, 
  D-01187  Dresden, Germany}

\title{Weak localization of light by cold atoms:
  the impact of quantum internal structure}

\date{\today}

\maketitle

\begin{abstract}
Since the
work of Anderson on localization, interference effects
      for the propagation of a wave in the presence of disorder have
      been extensively studied, as exemplified in coherent
      backscattering (CBS) of light. 
In the multiple scattering of light by a disordered sample of 
thermal atoms, interference effects are usually washed
out by the fast atomic motion. 
This is no longer true for cold atoms 
where CBS has recently been observed.
However, the
internal structure of the atoms strongly influences the 
interference properties.
In this paper, we consider light scattering by an 
atomic dipole  transition with arbitrary degeneracy 
and study its impact on  coherent backscattering.  
We show that the interference contrast is strongly reduced. 
Assuming a uniform statistical distribution over  
  internal degrees of freedom, we compute analytically the single and
  double scattering contributions to the intensity in the weak
  localization regime. The so-called ladder and crossed diagrams 
  are generalized to the case of atoms and permit
  to calculate enhancement factors and backscattering intensity
  profiles for polarized light and any closed atomic dipole
  transition.
\end{abstract}

\section{Introduction}

Interference of waves is the general feature shared by different
fields of physics such as Optics, Acoustics and Quantum Mechanics.
For waves propagating in disordered media, it was believed that
interference effects would be scrambled and that a reliable Boltzmann
transport theory would emerge~\cite{Chandrasekhar60}.  But
Anderson~\cite{Anderson58} predicted in the context of solid state
physics that interference can inhibit the propagation of matter waves
in disordered media (Anderson localization).  Since then, many
theoretical and experimental works have shown that elastic multiple
scattering in the presence of disorder is full of rich
phenomena~\cite{Rossum99,Berkovits94,Houches94}.  The coherent
backscattering effect, an interferential enhancement of the average
reflected light intensity in the backscattering direction, was the
first direct experimental evidence~\cite{Kuga84,vanAlbada85,Wolf85}
that interference of light waves persists in the presence of disorder
and has been extensively studied for the past fifteen years.

At the same time, considerable advances were achieved in creating and
controlling dilute gases of cold atoms, leading to the experimental
observation of Bose-Einstein condensation in 1995 and triggering
active experimental and theoretical research~\cite{BEC96}.  
It is not surprising that cold atomic gases
have been suggested as promising media for strong (Anderson)
localization of light~\cite{Nieuwenhuizen94}.
Well-defined atomic transition
lines allow strongly resonant light scattering
with cross-sections  of the order of the squared
optical wavelength, much bigger than the actual size of
the atom. 
In this respect, atoms are natural realizations of the
mathematical concept of point dipole scatterers
(also known as resonant Rayleigh scatterers), 
a paradigmatic model in the context of multiple 
scattering~\cite{Lagendijk96,deVries98,Orlowski00}.  
However, this simplified description has become questionable. 
The atomic dipole transition interacting with light in real
experiments is usually more complicated: both the
ground state (with angular momentum $J$) and the excited state (with angular
momentum $\Je$) present an important degeneracy, necessary for cooling
and trapping. 
This internal structure makes the atom behave very
differently from a point dipole scatterer.
Indeed, coherent backscattering of polarized light by a
laser-cooled gas of Rubidium atoms has been
observed recently~\cite{Labeyrie99,Labeyrie00} 
in the weak localization regime. 
There, surprisingly low enhancement factors for the backscattered
intensity indicate that interference is less
efficient for atoms than for classical point dipole scatterers.  
A careful study of the coherent propagation of light waves in atomic gases
therefore promises to be of great interest for both fields ``multiple
scattering in disordered media'' and ``cold atoms''.

In this 
paper, we show in detail how the internal atomic structure can
account for the reduction of the enhanced backscattering of polarized
light by atoms. 
In particular, we generalize the theory of single and
double scattering of polarized light by classical point scatterers to
the case of atomic scatterers with an arbitrarily 
degenerate dipole transition. 
Because of this degeneracy, the full atomic scattering tensor has to be
considered. It will be shown that its 
non-scalar parts are responsible for 
a single scattering background in all polarization channels and 
a drastic reduction of the interference contrast.

The paper is organized as follows.  Sec.~\ref{cbs} introduces the basic
notions of enhanced 
backscattering of light by a standard disordered medium. 
Sec.~\ref{amplitude.sec} presents an analysis of single and double
scattering amplitudes of light by atoms and shows qualitatively 
how a quantum internal
structure reduces the backscattering enhancement.  In
Sec.~\ref{analytical.sec}, the single and double scattering
intensities are calculated analytically, preparing the way for the
quantitative analysis contained in
Sec.~\ref{results.sec}.

\section{Enhanced backscattering of light}
\label{cbs}

\subsection{Two-wave interference}
A wave, characterized by its wavelength $\lambda$ or wavenumber
$k=2\pi/\lambda$ in vacuum, incident upon a disordered medium 
is scattered into a
multitude of partial waves. If the individual scattering events are
elastic, these partial waves are all coherent and interfere.  In the
weak localization regime, individual scatterers with a scattering
cross-section $\sigma$ are distributed with number density $n$ so that
the scattering mean free path $\ell=1/n\sigma$ is much larger than
$\lambda$. This condition, equivalently stated as $k\ell\gg1$, says 
that the mean distance between scattering events  
is much larger than the wavelength, so that waves propagate almost freely
inside the medium.   
In this regime,
 the wave amplitude $A$ can be constructed by coherent superposition
$A=\sum_p a_p$ of partial waves which are scattered along a
quasi-classical path joining the positions of consecutive
scatterers~\cite{Chakravarty86}.  The positions of all scatterers in
turn determine the precise shape of the resulting interference
pattern, as observed in the speckle figures of scattered laser light.

This interference pattern is naively expected to be washed out when
averaged over the realizations of the disorder (for example, by 
thermal motion of the scatterers).  In fact, the average intensity
$I=\mv{|A|^2}$ separates into independently squared amplitudes and the
sum of interference terms, $I=\sum_p \mv{|a_p|^2}+\sum_{p \ne p'}
\mv{\bar a_p a_{p'}}$ (the brackets indicate an average over
realizations of disorder, the bar denotes complex conjugation).  
If the scatterers are distributed
randomly, different scattering paths ($p'\ne p$) involve uncorrelated
phases.  The interference terms may be expected to vanish, $\mv{\bar
  a_p a_{p'}}=0$, yielding the uniform average intensity that is
familiar to us from the view of most natural objects like clouds.  In
the context of 
light scattering, 
it was first realized by 
Watson~\cite{Watson69}, de Wolf~\cite{deWolf71} and others, 
however, that
each multiple scattering sequence visiting $N$
scatterers in a given order $(1,\dots,N)$ has exactly one reversed counterpart
$(N,\dots,1)$. 
The phase difference between the two corresponding partial waves  
(visiting the same scatterers, but travelling in opposite directions)  
is given
by $\Delta\phi=({\bf k}+{\bf k}')\cdot({\bf r}_1-{\bf r}_N)$, where ${\bf k}$
and ${\bf k}'$ are the incoming and outgoing wave vectors, and ${\bf r}_1$
and ${\bf r}_N$ are the positions of the first and last scatterer along
the scattering path.  The phase difference is exactly zero in the
backscattering direction where ${\bf k}' = -{\bf k}$. Zero phase difference
means constructive interference, independently of the actual path
configuration.  This constructive two-wave interference therefore survives the
ensemble average and gives rise to coherent backscattering, the 
enhancement of the multiply scattered intensity in the
backward direction by a factor of two.

\subsection{Enhancement factor}  
There is an exception to the 
systematic interference
between direct and reverse amplitudes: scattering paths 
which are their own reversed do not give rise to 
any interference
term and thus add a uniform background 
to the average scattered intensity. 
In the weak scattering regime $k\ell\gg1$, 
this uniform background reduces to the single 
scattering contribution $I_{\mathrm S}$. 
In this regime, the average intensity can be
written as a sum of three terms
$I(\theta)=I_{\mathrm S}(\theta)+I_{\mathrm L}(\theta)+I_{\mathrm C}(\theta)$ 
as a
function of the angle $\theta$ with respect to the backscattering
direction.  Here, the so-called ladder term $I_{\mathrm L}(\theta)$ is the contribution
of all squared multiple scattering amplitudes, neglecting interference.  The
so-called crossed term $I_{\mathrm C}(\theta)$ contains the interferences  
between direct and reverse amplitudes.   
Under well chosen experimental
conditions, where all paths and their reverse counterparts have exactly the
same amplitude, the constructive two-wave interference leads to a 
maximal contrast $I_{\mathrm C}(0)=I_{\mathrm L}(0)$. 
Away from the backward direction, $I_{\mathrm C}(\theta)$ is averaged to zero
once the typical phase difference of interfering amplitudes approaches
$\Delta\phi\approx 1$.  Taking the double scattering contribution
($N=2$), the distance $r_{12}=||{\bf r}_1-{\bf r}_2||$ will be given on
average by the scattering mean free path $\ell$. To first order in
$\theta$, the typical phase difference then is
$\Delta\phi=k\ell\theta$. Therefore, $I_{\mathrm C}(\theta)$ decreases to zero
over an angular scale $1/k\ell$ which is very small in the weak
scattering regime $k\ell\gg1$.  Higher orders of scattering involve
paths with endpoints further apart and thus contribute to 
$I_{\mathrm C}$ with a 
smaller angular width.  For a semi-infinite and non-absorbing
scattering medium, the sum of all contributions has been shown to result
in the so-called coherent backscattering cone, a sharp intensity peak
exactly in the backscattering
direction~\cite{Akkermans88,vanderMark88,Gorodnichev90}.  When higher
orders of scattering become relevant, the width of the backscattering
enhancement is determined not by the scattering mean free path but
rather by the transport mean free path
$\ell_{\mathrm{tr}}=\ell/(1-\mv{\cos\theta})$; here, $\mv{.}$ denotes an
average  over the differential cross-section. If $\mv{\cos\theta}=0$, the
  two length scales are identical, $\ell_{\mathrm{tr}}=\ell$. This is true
for  
isotropic point scatterers and unpolarized atoms (cf.
  Sec.~\ref{single.sec}), so scattering and transport mean free path
  will be identified throughout the rest of this article. 
$I_{\mathrm S}$ and
  $I_{\mathrm L}$ exhibit a smooth angular dependence 
with respect to the
  normal of the surface of the medium (Lambert's
  law~\cite{BornWolf}). They can thus be taken constant, for not too oblique
  incidence, on the backscattering angular scale $1/k\ell$.
  
  The ratio of the average intensity at backscattering
  $I(0)=I_{\mathrm S}+I_{\mathrm L}+I_{\mathrm C}(0)$ to the average background intensity
  $I(k\ell\theta \gg 1)=I_{\mathrm S}+I_{\mathrm L}$ is the enhancement factor
\begin{equation}
\alpha=1+\frac{I_{\mathrm C}(0)}{I_{\mathrm S}+I_{\mathrm L}}. 
\label{efactor.eq}
\end{equation} 
Its maximum value $\alpha=2$ is attained if and only if  
there is
no single scattering background, $I_{\mathrm S}=0$, and the contrast
of the two-wave interference is perfect, $I_{\mathrm C}(0)=I_{\mathrm L}$.

\subsection{Polarization and reciprocity}

Since light is a vector wave, polarization (which 
describes the direction of the electric field vector) is an essential 
ingredient of any analysis of the enhancement factor. 
The incident field polarization $\beps$ and 
the scattered field polarization $\beps'$ define two
sets of orthogonal polarization channels. 
For linearly polarized incident light, the scattered light 
can be analyzed with parallel ($\lpar$) or perpendicular ($\lperp$) 
polarization. 
For circularly polarized incident light,
it is convenient to use the concept of helicity, i.e., 
the orientation of the circular polarization with respect
to the direction of propagation. 
The scattered light can be analyzed with preserved helicity 
($\hpar$) or flipped helicity ($\hperp$). 
At exact backscattering, these two cases respectively
correspond to flipped ($\beps'=\bbeps$) and
preserved polarization ($\beps'=\beps$).
Note that the circularly polarized light
scattered backwards by a mirror has the same polarization, 
thus flipped helicity.
      
  For classical scatterers, the following results have been established
~\cite{Mishchenko92,vanTiggelen97}:
$(i)$ Single
scattering in the backscattering direction is absent in the $\lperp$
and $\hpar$ channels for scatterers of spherical symmetry;
$(ii)$ In the absence of an external
magnetic field, the reciprocity theorem (see below) assures that 
$I_{\mathrm C}(0)=I_{\mathrm L}$ in the
parallel channels $\lpar$ and $\hpar$.  Satisfying simultaneously 
conditions $(i)$ and $(ii)$, an enhancement factor $\alpha=2$  
has been predicted and observed for spherically symmetric 
scatterers in the $\hpar$
polarization channel~\cite{Wiersma95}. 

 As reciprocity is an important notion for CBS, 
let us precise this point. 
Reciprocity is a symmetry property stemming from the
invariance of the fundamental microscopic dynamics under 
time-reversal~\cite{vanTiggelen97}.  Reciprocity assures 
that amplitudes relating to scattering
processes where initial and final states are exchanged and time-reversed
are equal. 
For the scattering of incident light with wavevector ${\bf k}$ and 
polarization $\beps$ into light with wavevector ${\bf k}'$ and 
polarization $\beps'$, it implies 
\begin{equation}
T_{\mathrm dir}({\mathbf k} \beps \rightarrow {\mathbf k}' \beps') =
T_{\mathrm rev}(-{\mathbf k}' {\bbeps'} \rightarrow -{\mathbf k} \bbeps).
\label{reciprocity_cl}
\end{equation}
Here,  $T_{\mathrm dir}({\bf k } \beps \rightarrow {\bf k}' \beps')$ is the  
amplitude of a given 
scattering sequence, and $T_{\mathrm rev}(-{\mathbf k}' {\bbeps'} \rightarrow
-{\mathbf k} \bbeps)$ is the amplitude of the reciprocal process  
(the bar denotes complex conjugation). 
In general, these reciprocal amplitudes describe different scattering 
processes and thus cannot interfere. CBS interference arises between 
amplitudes 
$T_{\mathrm dir,rev}({\bf k } \beps \rightarrow {\bf k}' \beps')$ 
associated to direct and  reverse scattering
paths 
with the same initial and final direction of propagation
and the same polarization.  
The reciprocity relation (\ref{reciprocity_cl}) 
thus assures equality for the two CBS amplitudes
if and only if two conditions are met: 
\begin{equation}
 {\bf k}' = - {\bf k} \quad \text{and} \quad  \bbeps' = \beps.
\label{recondition.eq}
\end{equation}
>From these conditions, it follows that the CBS amplitudes
of any given path and its reverse are equal 
at backscattering in the $\lpar$ and $\hpar$
channels, implying $I_{\mathrm C}(0) = I_{\mathrm L}$. 
On the other hand, 
away from the backscattering direction or in the perpendicular 
channels, the relation~(\ref{reciprocity_cl}) is still valid,
but says nothing about the pairs of amplitudes 
that interfere  
for CBS. These amplitudes are therefore expected to be  
different, leading to a
reduced contrast $I_{\mathrm C}< I_{\mathrm L}$.

\section{Amplitudes for scattering of light by atoms}
\label{amplitude.sec}

\subsection{Description of the atomic medium and approximations}

We are interested in the situation where the scatterers are not
macroscopic objects, but individual atoms. One may think of several
specific 
characteristics 
of atomic light scatterers which affect
coherent backscattering:

\begin{itemize}
\item Atoms have extremely \textit{narrow resonances}. 
Close to an atomic resonance,
the light scattering cross-section is of the
order of the square of the wavelength, much larger than
the geometric cross-section of the atom. 
A dense cloud of atoms therefore is ideal for strong elastic  
multiple scattering.

\item Because of the high polarizability of atoms near
an atomic resonance, it is rather easy to induce {\em non-linear}
effects (e.g. saturation) with only few milliwatts of laser power. 
Despite some studies of multiple scattering in non-linear 
media~\cite{Heiderich95}, it is basically unknown 
how this affects CBS by atoms.

\item  When an atom scatters a photon, its velocity
changes by an 
amount of the order of mm/s. This \textit{recoil} effect becomes important for 
cold atoms with typical velocities of a few cm/s. 

\item The atomic resonances being very narrow, atoms may be driven in
or out of resonance because of the \textit{Doppler} effect.
Adding the contributions of the various velocity classes to the
CBS signal is far from obvious. 

\item Atoms also have a \textit{quantum internal structure}. For a given
transition line, the total angular momentum $J$ of the atomic 
groundstate in general is not zero. 
In the absence of
any external magnetic field, the groundstate then is $(2J+1)$-fold degenerate.
As a first consequence, there
is the possibility of 
elastic light scattering processes
which change 
the internal atomic substate (degenerate Raman
transitions). Subsequent light scattering then gives rise to 
optical pumping.

\item When the atoms are very cold, their de Broglie wavelength 
becomes comparable to the optical wavelength. 
In this regime, the external atomic motion must be treated 
quantum mechanically. 
For high enough density, Bose-Einstein condensation sets
in.    
\end{itemize}

Addressing all these problems is beyond the scope of this paper.
We will focus our present investigation 
on the crucial role of the atomic internal structure, 
making use of several simplifying approximations.
 
Firstly, we assume the weak scattering relation
$k\ell\gg1$ to hold. This will be the case for sufficiently low 
density $n$ of the atomic
medium. Indeed, as the resonant atomic cross-section
$\sigma=1/n\ell$ is of the order of $\lambda^2$, weak scattering is
implied by the low density condition $n\lambda^3\ll 1$. 
In this regime, the independent
scattering approximation (ISA) is justified~\cite{Lagendijk96}. 
Equivalently, recurrent
scattering (i.e., sequences visiting a given scatterer more than once)   
can be neglected. The single scattering transition
matrix then suffices to compute the single scattering intensity which, 
in turn, serves as a building block for higher order scattering.
In this regime, the average index of refraction of the medium
is very close to unity (cf. Sec. \ref{effmedium.sec}). 

Secondly, we use quantum mechanical perturbation theory 
to describe the scattering of light by an atom. This will be valid as long 
as the laser intensity is sufficiently low~\cite{CCT}. 
We will restrict our calculation to the
case of one photon scattering, determining the transient reponse
of the system rather than its stationary state. 
This method thus ignores saturation
effects and optical pumping. In principle both could be described by
carrying the perturbation to higher numbers of scattered photons, but
practically one has to calculate the stationary density matrix by
solving the multi-level optical Bloch equations. In the experimental
application sofar \cite{Labeyrie00}, the laser intensity was kept 
well below the saturation
intensity. Furthermore, optical pumping in the bulk of an optically 
thick atomic cloud is expected to be severely limited 
by multiple scattering.

Thirdly, we treat the 
external motion of the atoms classically. In other words, we require
the atoms to be sufficiently hot so that the coherence length
of the external wavefunction is shorter than the optical wavelength. 
This is the case for cold atoms created in a standard magneto-optical
trap.  The present treatment does not apply, however, to ultra-cold atoms as,
e.g., in a Bose-Einstein condensate. 

The question of the recoil effect can then be addressed rather
easily. Indeed, light imparts various momentum kicks to the 
atoms defining a scattering path, but these momentum transfers are 
identical for the direct and reverse paths at backscattering. 
Consequently, the recoil effect does not affect the interference between
the amplitudes along the two paths.

Let us now discuss the role of the atomic motion. 
Close to an atomic resonance of width $\Gamma$, the average time 
an atom takes to scatter a photon is $\Gamma^{-1}$. If, in the 
meantime, atoms move by more than an optical wavelength, then
the interference term between direct and reverse scattering sequences 
will be spoiled \cite{Golubentsev84}. 
To avoid this, we require the spread 
$v$ of the atomic velocity distribution to satisfy
\begin{equation}
  \label{eq:doppler}
  kv \ll \Gamma.
\end{equation}
If this condition is met, atoms can be thought as being
fixed in space during the multiple scattering process
~\cite{Labeyrie00}. 
On a much longer
time scale, the motion of the atoms simply acts as a configuration 
average. Typically, 
eq.~(\ref{eq:doppler}) is satisfied for atoms slower
than few m/s, which is true for atoms originating from
a magneto-optical trap.
Note that eq.~(\ref{eq:doppler}) can be alternatively 
viewed as a resonance condition: under scattering, the Doppler shift 
will not bring atoms out of resonance. 
  
\begin{figure}
  \centering \includegraphics[width=0.40\textwidth]{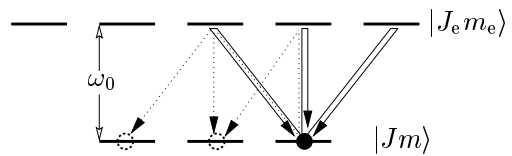}
\caption{Energy representation of a degenerate atomic dipole
  transition, here for $J=1$, $\Je=2$.  Arrows mark atomic transitions
  from the initial substate, here $\ket{J,m=J}$, under scattering 
  of a photon. In the absence of an external magnetic
  field, all transitions are elastic. 
  Solid arrows: Rayleigh transitions, conserving
  the magnetic quantum number ($m'=m$).  Dotted arrows: degenerate
  Raman transitions, changing the magnetic quantum number (here
  $m'=m-1$ and $m'=m-2$). }
\label{transition.fig}
\end{figure}

  Under these conditions, the most important
effect will come from the internal structure of the atoms,
i.e., the degeneracy of the light scattering transition. 
We assume that the incident light
is nearly resonant with an atomic
transition of (bare) angular frequency $\omega_0$ between a ground
state with total angular momentum $J$ and an excited state with total angular
momentum $\Je$ (see fig.~\ref{transition.fig}). 
Since no external magnetic field is supposed to be present,
the atomic ground state and the excited state are 
respectively $(2J+1)$-fold and $(2\Je +1)$-fold degenerate. 
The corresponding substates with magnetic quantum numbers $m$ and
$\me$ with respect to some arbitrary quantization axis 
are denoted by $\ket{Jm}$ for the
ground state ($-J\le m\le J$) and by $\ket{\Je \me}$ for the excited state 
($-\Je\le \me\le \Je$). 

The restriction to a single $J\to \Je$ transition could be
relaxed at the price of more complicated calculations since, in essence, 
the various transitions contribute
independently to the atomic scattering tensor, the essential ingredient 
of our analysis as shown below.
We will also assume 
that the  $J\to \Je$ transition is closed, 
so that light 
scattering is purely elastic. Again, different final states 
with different energies could be included 
along the same lines of reasoning.

In the following, we recall the amplitude for the scattering of one
photon by one atom (single scattering) and determine the amplitude for
the scattering by two atoms (double scattering). We discuss 
how the degeneracy of the atomic
dipole transition affects the CBS enhancement factor (\ref{efactor.eq}).  
We will use a full quantum mechanical treatment
of both atoms and electromagnetic field. 
While the internal atomic degrees of freedom 
can only be described 
quantum mechanically, the 
electromagnetic field 
will be described  by quantum Fock states 
for reasons of symmetry. 
An equivalent treatment can be set up for low-intensity coherent states which
are known to correspond closely to a classical light
field~\cite{Loudon73}. 

Throughout the paper, transitions between identical atomic substates 
($m'=m$) are called \emph{Rayleigh transitions} and 
transitions between different substates
($m' \ne m$) are called \emph{degenerate Raman transitions}. 
Let us stress that, since one-photon scattering on a degenerate 
atomic dipole transition is necessarily elastic, 
inelastic processes (also known as Raman \emph{scattering}) are completely 
absent of our analysis.
In the following, we use natural units where $\hbar=c=1$ so that
$[\mbox{length}]=[\mbox{time}]=[\mbox{frequency}]^{-1}
=[\mbox{energy}]^{-1}$. 

\subsection{Single scattering amplitude}
\label{singleamp.sec}

In the single scattering situation, an atom at fixed position ${\bf r}$
is exposed to a plane light wave with wave vector ${\bf k}$, angular
frequency $\omega=k$ and transverse polarization $\beps$. 
We describe the
uncoupled system by the sum of the atomic internal hamiltonian and the
free field hamiltonian,
\begin{equation}
H_0=\omega_0\sum_{\me}\ket{\Je \me}\bra{\Je \me} 
+ \sum_{{\bf k}, \beps \perp {\bf k}} \omega \: 
	a_{{\bf k}\beps}^\dagger a_{{\bf k}\beps}.
\label{h0.eq}
\end{equation} 
Here, $a_{{\bf k}\beps}$ and $a_{{\bf k}\beps}^\dagger$ are the usual
annihilation and creation operator of a transverse field mode with
wave vector ${\bf k}$ and polarization vector $\beps$.
The corresponding one-photon Fock state will be denoted  
$\ket{{\bf k}\beps}$ where the transversality
$\ps{k}{\beps}=0$ is understood.  
The interaction between
atom and light is given in the dipole form by 
$V=-{\bf D}\cdot {\bf E}({\bf r})$.  
The atomic dipole operator ${\bf D}$ connects the subspaces
${\mathcal H}_J$ and ${\mathcal H}_{\Je}$ (since we consider a closed
transition, no other subspaces are involved) with reduced matrix
element $\bra{\Je}|{\bf D}|\ket{J}=D\sqrt{2\Je+1}$ \cite{Edmonds60}.  
The electric field
operator at the atomic position is given by
\begin{equation}
{\bf E}({\bf r})= i \sum_{{\bf k}, \beps \perp {\bf k}}{\mathcal E}_\omega 
\beps_{{\bf k}\beps} a_{{\bf k}\beps}\exp[i({\bf k}\cdot {\bf r})]+h.c.
\end{equation}
The field strength ${\mathcal E}_\omega=(\omega/2\epsilon_{0} L^3)^{1/2}$ is
defined in terms of a quantization volume $L^3$ that eventually disappears in
results of physical significance.

The probability amplitude for a transition from an initial state
$\ket{i}=\ket{{\bf k}\beps;Jm}$ to a final state 
$\ket{f}=\ket{{\bf k}'\beps';Jm'}$ is the element $S_{fi}$ of the scattering
matrix. The 
transition amplitude for $i\ne f$ 
is written 
\begin{equation}
S_{fi}=-2i \pi \delta(\omega-\omega')
T_{fi}(\omega+i0),
\end{equation} 
in terms of the transition operator $T$. 
Here, because the atomic ground state is degenerate, 
energy conservation,  
assured by the delta distribution, 
implies elastic light scattering $(\omega'=\omega)$.
The matrix element $T_{fi}(z)$ is calculated perturbatively
using the Born expansion $T(z)=V+VG_0(z)V+\dots$ in
powers of the interaction 
$V$ 
and the resolvent $G_0(z)=(z-H_0)^{-1}$ of
the  unperturbed system. 
The excited atomic state can be eliminated by
partial summation of the Born series, dressing the transition
frequency and introducing a finite lifetime~\cite{CCT}:
$\delta=\omega-\omega_0$ is the detuning from the (dressed) transition
frequency $\tilde\omega_0\approx \omega_0$ and $\Gamma=D^2
\omega_0^3/3\pi\epsilon_{0}$ is the natural width of the atomic
excited state. 

\begin{figure}
  \centering \includegraphics[width=0.35\textwidth]{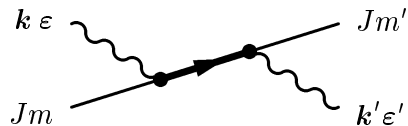}
\caption{Feynman diagram for the transition matrix
  element $T_{fi}$ (eq.~(\ref{singleamp.eq}))
  near resonance.  Wavy lines denote photons, straight lines atomic
  internal states.  The thick line stands for
  the 
  dressed 
  propagator $(\delta+i\Gamma/2)^{-1}$ of the excited atomic
  state.  The absorption vertex yields a factor 
 $\omega_{\mathrm R} \exp(i{\bf k}\cdot {\bf r})
 \bra{\Je \me}\beps\cdot {\bf d}\ket{Jm}$, the emission vertex
  yields a factor 
$\omega_{\mathrm R} \exp(-i{\bf k}'\cdot {\bf r})
\bra{Jm'}\bbeps'\cdot {\bf d}\ket{\Je \me}$, 
and all intermediate variables, here $\me$, have to be summed over.
  }
\label{singleamp.fig}
\end{figure}

Let us define the reduced dipole operator 
${\bf d}={\bf D}/D$ and the Rabi frequency 
$\omega_{\mathrm R}= D {\mathcal E}_\omega$. 
The transition matrix element  
$T_{fi}=\bra{{\bf k}'\beps',Jm'}T(\omega+i0)\ket{{\bf k}\beps,Jm}$ 
near resonance 
then is
\begin{equation}
T_{fi}
=  
\frac{\omega^2_{\mathrm R}}{\delta+i\Gamma/2} \bra{Jm'}(\bbeps'\cdot {\bf d})
 (\beps \cdot {\bf d})\ket{Jm}
\; e^{i({\bf k}-{\bf k}')\cdot {\bf r}} ,  
\label{singleamp.eq}
\end{equation}  
represented by its Feynman diagram in fig.~\ref{singleamp.fig}.  The
condition ``near resonance'' means $\delta\ll \omega_0$ (but not
necessarily $\delta<\Gamma$). Therefore, anti-resonant scattering,
i.e., first emission then absorption, can be neglected.

In eq.~(\ref{singleamp.eq}),
all information about the atomic internal degrees of
freedom and polarization has been factorized into the matrix element
\begin{equation} 
\bra{Jm'}\bbeps'\cdot {\bf t} 
\cdot \beps \ket{Jm} = 
\frac{\omega^2_{\mathrm{R}}}{\delta+i\Gamma/2}\bra{Jm'}(\bbeps'\cdot {\bf d})
(\beps \cdot {\bf d}) \ket{Jm}.
\label{scattensor.eq} 
\end{equation}
This defines the scattering operator ${\bf t}$ which acts on the product
space 
${\mathcal H}_J\otimes \C^3$ 
of atomic internal states and
polarizations.  It is the scattering operator ${\bf t}$ that
characterizes the scattering object and contains all relevant
information about the scattering process~\cite{LandauQED}.  
We can separate its frequency dependence from its tensor structure:
${\bf t}(\omega)= t(\omega) {\bf \hat t}$
where 
\begin{equation}
t(\omega) = \frac{\omega^2_{\mathrm{R}}}{\delta+i\Gamma/2}
\end{equation} 
is given as the ratio of the squared Rabi frequency 
(or squared coupling strength) and the resonant denominator known
for point dipole scatterers~\cite{Lagendijk96}.   
The novelty of the present approach lies in the peculiar 
tensor part ${\bf \hat t}$.
For a given transition $m\rightarrow m'$, the dimensionless matrix element 
\begin{equation}
 \hat{t}_{ij}(m,m') = \bra{Jm'} d_i d_j \ket{Jm}
\label{tmatrix.eq}
\end{equation}
defines the scattering tensor 
which connects the incoming to the outgoing polarization. 
This $3 \times 3$ $t$-matrix  
can be decomposed into its scalar, antisymmetric and symmetric
traceless components, transforming irreducibly under rotations 
\cite{Edmonds60}.

A classical point dipole scatterer is characterized by a $t$-matrix 
proportional to unity~\cite{Lagendijk96}.  
This behaviour is reproduced by the
elementary dipole transition $J=0$, $\Je=1$. Indeed, the only matrix
element of the scattering operator yields the scalar part 
$\bra{00}\hat t_{ij}\ket{00}=\delta_{ij}$. 
Non-spherical classical scatterers also display 
an additional traceless symmetric part in their scattering $t$-matrix. 
In the case of atoms,
therefore, it is the antisymmetric part that is characteristic 
for the quantum internal structure. 
The antisymmetric part simply implies that an atom 
scatters light with polarization-dependent strength. 
To be more specific, consider scattering
of circularly polarized light in a Rayleigh transition 
($m'=m$; quantization axis is the direction of propagation). The
scattering strengths for the two possible helicities are different because 
the Clebsch-Gordan coefficients associated to the transitions
$\ket{J,m} \leftrightarrow \ket{\Je,m\pm 1}$ are unequal. 

This situation is somewhat similar to the usual Faraday effect 
where circular polarizations with opposite helicities are scattered 
differently 
in the presence 
of an applied magnetic field \cite{Barron82}. There are, however,
significant differences:
in the Faraday effect, the antisymmetric part of the atomic polarizability
depends both on the magnetic field direction and on the direction
of light propagation; for the atomic scattering operator, antisymmetry is a 
fully intrinsic property.
When averaged over the internal state, the antisymmetric part
of the atomic scattering operator  vanishes,  
leading to a symmetric polarizability and to no 
dichroism inside the effective medium (cf. Sec.~\ref{effmedium.sec}). 
Thus, the degenerate atomic situation ressembles
a (zero magnetic field) Faraday effect depending on the internal state
of the atom.
 
The degeneracy of the atomic ground state also implies that, 
by scattering a photon, the
internal state may change (cf. Fig.~\ref{transition.fig}). 
The possibility of changing the
internal state leaves more choice for the photon polarization. Which
polarization is possible for which transition follows from the
conservation of angular momentum.  In exactly the backscattering
direction (${\bf k}'=-{\bf k}$) and chosing the quantization axis
along the direction of propagation, the following relations hold: 
linearly polarized light is scattered into the $\lpar$
channel by a Rayleigh transition ($m' = m$), and into 
the $\lperp$ channel by a degenerate Raman transition ($|m'-m|=1$); 
circularly polarized light is backscattered into the $\hperp$ channel by 
Rayleigh transition ($m' = m$), and into the $\hpar$ channel by 
a degenerate Raman transition ($|m'-m|=2$).

Therefore, the single scattering amplitude shows that changes in the
atomic internal state permit changes in the light polarization.  Since
in general the atomic internal state is not under control, the single
backscattering contribution cannot be removed by polarization analysis
(with the only exception $J=\frac{1}{2}$ in the $\hpar$ channel)
and degrades the observable enhancement factor
(\ref{efactor.eq}). 

\subsection{Double scattering amplitudes}
\label{doubleamp.sec}

\subsubsection{Direct and reverse transition amplitudes}
\label{direvamplitudes.sec}
 
\begin{figure}
  \centering \includegraphics[width=0.45\textwidth]{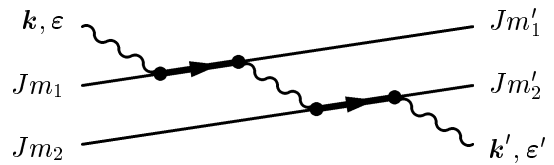}
\caption{Feynman diagram of the direct transition matrix element 
  $T_{fi}^{(\mathrm dir)}$ 
(eq.~(\ref{Tdir.eq})): 
resonant scattering first by atom $1$,
  then by atom $2$. 
The Feynman rules are defined in
  fig.~\ref{singleamp.fig}.}
\label{doubleamp.fig}
\end{figure}

In the double scattering situation, a plane wave impinges upon two
atoms $\alpha=1,2$ at fixed positions ${\bf r}_\alpha$. 
The interaction between atoms and field in the full hamiltonian is now 
$V = -{\bf D}_1\cdot {\bf E}({\bf r}_1)
-{\bf D}_2\cdot {\bf E}({\bf r}_2)$. 
This interaction defines  a transition operator
$T$ along the lines of Sec.~\ref{singleamp.sec}. 
Resonant dipole interaction between the atoms arises from the
exchange of photons.  
Among the
numerous different diagrams that describe the transition
$\ket{i}=\ket{{\bf k}\beps,Jm_1,Jm_2}\rightarrow \ket{f}= \ket{{\bf k}' 
 \beps',Jm'_1,Jm'_2}$, the two dominant diagrams involving both atoms
are concatenations of two single scattering diagrams.  The first
diagram, shown in fig.~\ref{doubleamp.fig}, describes the direct
scattering path: absorption of the incident photon by atom 1 and
emission of the final photon by atom 2.  
The diagram for the reversed
path is obtained by exchanging the role of the two atoms. 

The Feynman rules introduced in fig.~\ref{singleamp.fig} permit to
explicit the scattering amplitudes.  As usual in diagrammatic
expansions, a sum over the virtual intermediate states has to be
carried out, here the excited atomic states and the 
intermediate photon.  
The matrix element for the direct
scattering path in the far field
approximation $kr_{12}\gg1$ takes the following form: 
\begin{equation}
T^{(\mathrm dir)}_{fi}
= - \frac{3 \Gamma t^2(\omega)}{4 \omega^2_{\mathrm R}}\,
\frac{\exp(ik r_{12})}{k r_{12}}\, 
\hat t_{\mathrm dir}\,
e^{i({\bf k}\cdot {\bf r}_1- {\bf k}'\cdot {\bf r}_2)} .
\label{Tdir.eq}
\end{equation}
Again, all information about polarization and internal structure is
factorized into the dimensionless matrix element
\begin{equation}
\hat t_{\mathrm dir}
=\bbeps'\cdot {\bf \hat t}_2(m_2,m_2') \cdot  
    \Delta\cdot{\bf \hat t}_1(m_1,m_1') \cdot \beps.
\label{internaldir.eq}
\end{equation}
Here, the dimensionless one-atom 
$t$-matrices, defined in
eq.~(\ref{tmatrix.eq}), are connected by the projector
$\Delta_{ij}=\delta_{ij}-\hat n_i\hat n_j$ onto the plane transverse
to the unit vector ${\bf\hat n}={\bf r}_{12}/r_{12}$ joining the two
atoms.
 
The matrix element $T^{(\mathrm rev)}_{fi}$ for the reversed
path is obtained from eq.~(\ref{Tdir.eq}) by exchanging the roles of atoms 1
and 2.  The internal matrix element 
(\ref{internaldir.eq}) becomes  
  \begin{equation}
\hat t_{\mathrm rev}
=\bbeps'\cdot {\bf \hat t}_1(m_1,m_1') \cdot 
    \Delta\cdot{\bf \hat t}_2(m_2,m_2') \cdot\beps  . 
\label{internalrev.eq}
\end{equation}

The two amplitudes
$T^{(\mathrm dir)}_{fi}$ and $T^{(\mathrm rev)}_{fi}$ describe
indistinguishable processes and interfere. 
A maximal contrast in the backscattering direction is
obtained if and only if the amplitudes have equal magnitude $\hat t_{\mathrm
  dir}=\hat t_{\mathrm rev}$.   
But due to the 
non-scalar part of the atomic $t$-matrix, we expect that in general
the matrices do not commute,  
$\bs{\hat t}_2 \cdot \Delta \cdot \bs{\hat t}_1
\ne \bs{\hat t}_1 \cdot \Delta \cdot \bs{\hat t}_2$, so that 
(\ref{internaldir.eq}) and (\ref{internalrev.eq}) are not equal. 
An exception to this rule is of course the case $m_1=m_2$, $m_1'=m_2'$
where the exchange symmetry assures their equality.  
Furthermore, 
we can see that it is precisely
the antisymmetric part of the $t$-matrix that is responsible for the
inquality of amplitudes in the parallel polarization channels.  
Indeed, if the one-atom $t$-matrix were symmetric,  then 
$\bbeps'\cdot {\bf \hat t}_1(m_1,m_1')\cdot \Delta \cdot 
	{\bf \hat t}_2(m_2,m_2')  \cdot \beps 
	=\beps\cdot {\bf \hat t}_2(m_2,m_2')\cdot \Delta \cdot 
	{\bf \hat t}_1(m_1,m_1')\cdot \bbeps'$. 
In the parallel channels $\bbeps' = \beps$,
from this would immediately follow the equality of 
the direct and reverse matrix elements (\ref{internaldir.eq}) and
(\ref{internalrev.eq}).  
But because of the antisymmetric part of the atomic scattering tensor, 
in general 
\begin{equation}
\begin{split}   
\bbeps' \cdot 
{\bf \hat t}_1(m_1,m_1') & \cdot  \Delta 
	\cdot {\bf\hat t}_2(m_2,m_2') \cdot \beps  \ne   
\\ & 
 \beps \cdot  {\bf\hat t}_2(m_2,m_2')  \cdot \Delta 
	\cdot{\bf\hat  t}_1(m_1,m_1') \cdot\bbeps', 
\end{split}
\end{equation}
so that the two interfering amplitudes are different in magnitude.  
An explicit example for unequal interfering amplitudes
in the $\hpar$ channel 
-- one is zero while the other is not -- has been given 
in~\cite{Jonckheere00}.
  
\subsubsection{Reciprocity revisited}
\label{reciprocity.sec}

A question may arise at this point: Does the imbalance of amplitudes
$\hat t_{\mathrm dir}\ne \hat t_{\mathrm rev}$ 
contradict the theorem of reciprocity?  The
answer is no: the complete system ``field and atoms''
obeys
reciprocity, but this does not imply 
$\hat t_{\mathrm dir} = \hat t_{\mathrm  rev}$. The classical reciprocity 
relation~(\ref{reciprocity_cl}) has to be generalized to
take into account the set $\{m\}$ of the internal variables of 
all atoms~\cite{Landau}: 
\begin{equation}
\begin{split} 
T_{\mathrm dir}({\mathbf k}\beps,\{m\}\rightarrow  {\mathbf
k}'\beps',\{m'\}) = (-1) ^ {\sum_i(m_i' - m_i)} 
\\ \times 
  T_{\mathrm rev}(-{\mathbf k}' {\bbeps'},-\{m'\}\rightarrow -{\mathbf k}
\bbeps,-\{m\}).
\label{reciprocity_quant}
\end{split}
\end{equation}
This relation shows that in order to obtain the reciprocal sequence 
of a given sequence, the signs of all internal 
quantum numbers have to be flipped. 
The reciprocity relation (\ref{reciprocity_quant}) assures the equality of 
two interfering CBS amplitudes only if three conditions are met: the two
classical conditions (\ref{recondition.eq}) on light 
direction and polarization, and a third one 
pertaining to the atomic internal variables,
\begin{equation}
\{m'\}=\{-m\}.
\label{atcondition.eq}
\end{equation}
Whereas the direction of observation and polarization can be controlled  
experimentally, this 
is impossible for the internal atomic states in
an optically dense medium. 
Just as in the case of scattering away from the backward direction or into 
perpendicular polarization channels, 
reciprocity including the internal states never ceases
to be valid, but simply becomes inapplicable to predict the equality of
interfering amplitudes.   
It follows that although there might be some
amplitudes satisfying condition (\ref{atcondition.eq}), the majority will not,
and perfect interference contrast is lost. 
Of course, in the case of the elementary 
dipole transition $J=0$, $\Je =1$, the condition (\ref{atcondition.eq}) is
trivially fulfilled since all atoms verify $m'=m=0$ and we recover the 
classical case.

Classical reciprocity 
for light scattering has been derived
using Maxwell's equations for a linear scattering medium provided that its
constitutive tensors (dielectric constant, permeablility and conductivity) be
symmetric~\cite{Saxon55}. In the present case,  
when one does not consider the internal atomic states
as intrinsic variables of the system but as given parameters for each path,  
the $t$-matrix  ${\bf\hat t}(m,m')$ 
then has an antisymmetric part. In this respect, atoms with
degenerate transitions constitute a scattering medium that does not
obey classical reciprocity, 
and a reduced interference is no surprise.
Indeed, the same is observed in scattering media with the Faraday
effect~\cite{MacKintosh88,Martinez94} where the external 
magnetic field is said to break time-reversal invariance.

Finally, let us note that the ensemble average over the internal
variables $\{ m \}$ cannot restore the equality of the direct and
interference contributions to the diffuse intensity. 
Indeed, $I_{\mathrm C}(0)$ is equal
to $I_{\mathrm L}$ if and only if, for each pair of 
scattering paths, the
direct and reverse amplitudes are equal.  
In the sum of all contributions, the equal amplitudes cannot win back 
what the non-equal amplitudes have lost: the result is a reduced overall
enhancement factor.  

\subsubsection{The role of degenerate Raman transitions}
\label{Raman.sec}

In the case of the
elementary  dipole transition $J=0$, $\Je=1$, the atomic scattering
tensor only has a scalar part 
$\bra{00}\hat t_{ij}\ket{00}= \delta_{ij}$.  
The analysis of the double scattering amplitudes in 
Sec.~\ref{direvamplitudes.sec}
shows that the internal amplitudes then are equal and full
interference contrast is guaranteed. As for $J=0$ no
degenerate Raman transitions ($m'\ne m$) can occur, the following
question is inevitable: 
can the decrease of interference contrast be attributed 
to the Raman transitions alone?

Indeed, one might be tempted to suggest 
incoherence of the Raman scattered light (i.e., the loss of
phase coherence in spontaneous emission) as the origin 
for this loss of contrast. In the present description,
however, this is \textit{not} a pertinent explanation. It is true that the
Raman scattered light does not interfere with the reference light from
the source: the respective final atomic states are orthogonal 
and the two amplitudes do not describe indistinguishable processes  
(this is a typical ``which-path'' argument~\cite{Scully82,Itano98}). 
But a photon scattered elastically along the direct path interferes
very well with the same photon scattered along the reverse path --- as long
as the internal states of all atoms in both processes are identical, no matter
whether they describe degenerate Raman or Rayleigh transitions. 

A closer analysis of the situation in the channels of circular
polarization permits the following remarks. In the channel $\hperp$
of flipped helicity,  
a selection rule special to the double scattering configuration
admits only Rayleigh transitions to the crossed intensity. 
The ladder intensity contains a contribution from Rayleigh transitions
(equal to the crossed intensity) and an additional contribution from
degenerate Raman transitions. In
this sense, the degenerate Raman transitions are responsible for a
reduced double scattering interference in the $\hperp$ channel. 
This is consistent with the observation that 
degenerate Raman transitions make the atoms behave as 
non-spherical scatterers for which reduced interference in the
perpendicular channels is expected.  
But for higher scattering orders, Raman
transitions contribute also to the crossed intensity, and it is no
longer evident to compare the relative weights of Rayleigh and
Raman contributions.    

The explicit example of a double Rayleigh transition 
in the $\hpar$ channel 
with zero interference
given in ref.~\cite{Jonckheere00}, shows that Rayleigh
transitions also are responsible for a loss of contrast. 
On the other hand, 
the fact that Raman transitions give to atoms some characteristics
of non-spherical scatterers does not by itself imply a loss of
contrast: the reciprocity theorem is independent of the
actual shape of the scatterers and applies to spherical as well
as to non-spherical classical scatterers.
For example,
a double Raman transition such that
($m_1=m_2=-m_1'=-m_2'\neq 0$) satisfies the
reciprocity condition (\ref{atcondition.eq}) 
and has perfect contrast (as is also evident from the exchange
symmetry). 
In the sum of all scattering amplitudes, Raman
scattering amplitudes can even be dominant in the backscattering
interference signal. An explicit example of such a situation is given in
fig.~\ref{cone34.fig}. 

In fact, independently of the scattering order,  
it is precisely the antisymmetric
part of the atomic scattering tensor that is responsible for the loss
of contrast in the parallel polarization channels. This antisymmetric
part appears for both degenerate Raman and Rayleigh transitions
(cf. the unequal scattering of circularly polarized light with different 
helicities as mentioned in Sec.~\ref{singleamp.sec}). Therefore, the
degenerate Raman transitions must \textit{not} be held responsible
alone for the reduction of interference contrast.

\section{Analytical formulation of the internal ensemble average}
\label{analytical.sec}

We wish to describe the light propagation inside a macroscopic 
disordered medium on
average. Starting from an entirely symmetric microscopic 
description of matter and light,  
the ensemble average is a trace over the matter degrees of freedom.  
This trace contains 
an average over atomic positions as well as an
average over the internal degrees of freedom.  
This is analogous to
the case of classical non-spherical scatterers where averages over
position and orientation have to be performed. 
We suppose in the
following that the atomic sample is prepared without correlations
between positions and internal 
substates, and that different atoms
are uncorrelated. This is 
a reasonable assumption for 
a cloud of cold atoms created from 
a standard magneto-optical trap.  The two averaging procedures
then become independent. As far as positions are concerned, we will
use the averaging techniques developed for classical point
scatterers~\cite{Rossum99}. In the independent scattering approximation, 
the average over the internal quantum numbers $\{m\}$
can be expressed as traces over a one-atom density
matrix $\rho$ and one-atom operators.

\subsection{Average amplitudes: effective medium}
\label{effmedium.sec}

Tracing over the matter degrees of freedom defines an 
effective medium for the average propagation of light 
amplitudes and intensities.  
In this paragraph, we will deal with the
rather simple issue of the average amplitude. As will be seen, the internal
structure of the atomic scatterers provides no major surprise, and we are able
to recover the well-known properties of a dilute atomic gas~\cite{Loudon73}. 
The impact of the effective medium on the amplitude 
is described by the self-energy $\Sigma(\omega)$ that
renormalizes the vacuum light frequency $\omega$~\cite{Lagendijk96}. 
In the independent
scattering approximation, the self-energy is proportional to the average 
scattering operator, 
\begin{equation}
\bs \Sigma(\omega)= N \Tr \rho {\bf t}(\omega) =  
N \mv{{\mathbf t}(\omega)}_{\mathrm int}.
\end{equation}
Because of the vector character of the light wave, the self-energy 
is formally a second rank tensor.       
Assuming a scalar density matrix, i.e., a 
uniform distribution over internal states, 
the internal average simply projects onto the scalar part: 
$\bs \Sigma(\omega) = \Sigma(\omega) \bs 1$. 
Calculating the average is elementary using the closure relation of  
Clebsch-Gordan coefficients, and we find 
\begin{equation}
\Sigma(\omega) = n M_J \frac{3\pi}{\omega^2} \frac{\Gamma/2}{\delta+i\Gamma/2}
\end{equation} 
Here, we define for convenience 
the ratio of multiplicities 
\begin{equation}
  M_J=\frac{2\Je+1}{3(2J+1)}
\label{mj.eq}
\end{equation}
with $M_0=1$.  
The atomic polarizability close to resonance is given by 
\begin{equation}
\bs \alpha(\omega) = - \frac{2L^3}{\omega} 
\mv{{\mathbf t}(\omega)}_{\mathrm int} .
\label{polarizability.eq} 
\end{equation}  
Expliciting the internal average 
as a weighted sum over substates, 
\begin{equation}
 \mv{{\bf t}(\omega)}_{\mathrm int} 
   = \sum_m p_m\bra{Jm}{\bf t}(\omega)\ket{Jm},
\end{equation}
it is evident that solely the Rayleigh transitions $(m'=m)$ enter into the
definition of the self-energy and of the polarizability.  
In the case of a uniform distribution with weights $p_m=(2J+1)^{-1}$,  
this average selects the scalar part of the
scattering operator. 
A thorough discussion of the polarizability, the
scattering operator and its analysis by decomposition in irreducible
components 
can be found in the textbook
by Berestetskii et al.~\cite{LandauQED}. 

The susceptibility of the dilute atomic medium is $\chi= n\alpha$, and 
the condition of low density now reads $n\alpha\ll 1$. 
The effective refractive index then is given by 
$n_{\mathrm r}=1+n\alpha/2$. 
Its real part is very close to unity, 
and we need not distinguish between 
the optical wavelength in the medium and in the 
vacuum.  
Its imaginary part describes attenuation of the average amplitude, and here the effect
of the dilute medium is essential. Since we do not describe any
absorption, all attenuation is necessarily due to scattering from the initial
mode into other modes. This argument is the essence of the optical theorem 
\begin{equation}
\sigma_{\mathrm tot} = - 2 L^3 {\mathrm Im} 
     \mv{\bbeps \cdot {\bf t}(\omega) \cdot\beps}_{\mathrm int} 
 = k \ {\mathrm Im}  
\alpha(\omega).
\end{equation}     
We thus find the total scattering cross-section  
\begin{equation}
 \sigma_{\mathrm tot} = M_J  
 \frac{6\pi}{k^2}\frac{1}{1+4\delta^2/\Gamma^2} .  
\label{crossection.eq}
\end{equation}
This well known 
expression features the resonant dipole cross-section
$6\pi/k^2=3\lambda^2/2\pi$ and the Lorentzian line shape for detuning
$\delta$ around the resonance with width $\Gamma$. 

The mean free path of the light inside
the average medium is $\ell= -2({\mathrm Im} \Sigma(\omega))^{-1}$.  
By virtue of the optical theorem, 
it depends on the total cross-section and on the number density
of scatterers through:
\begin{equation}
\ell = \frac{1}{n\sigma_{\mathrm tot}}
\end{equation}
and is independent of both the polarization and the direction
of propagation. This reflects statistical invariance of the 
atomic medium under rotation. 

In summary, in the weak density and weak scattering regime, the internal
structure has very small influence on the properties of an average light 
amplitude. For a uniform statistical distibution over internal states, 
all average
quantities are isotropic and are only modified by a factor
$M_J=(2\Je+1)/3(2J+1)$ with respect to the classical dipole point
scatterer where $M_0=1$. 
This is not surprising since the internal average over a scalar
density matrix simply selects the scalar part of the atomic scattering
operator.  The antisymmetric part that appeared as the genuine quantum
feature in Sec.~\ref{amplitude.sec} therefore is not present here. 

\subsection{Average intensities}
\label{ensembleav.sec}

Coherent backscattering is an interference
effect for the average intensity, which of course must be distinguished from
the square of the average amplitude.  
Consequently, we stress that it is not sufficient to calculate 
quantities pertaining to the average amplitude (such as the polarizability or
the scattering mean free path) in order to decide whether the internal
structure affects CBS or localization. 
In the following paragraphs, we show how proper use of tensor algebra 
makes it possible to analytically 
perform the averaging over the atomic internal degrees of freedom.
In the specific case of a semi-infinite medium, exact 
averaging over the external position of the atoms is also 
possible for the single and double scattering contributions.
     
The average scattered intensity $I$ far away from the medium can be
calculated in terms of the dimensionless bistatic
coefficient~\cite{Ishimaru78}
\begin{equation}
\gamma=\frac{4\pi}{A}
\mv{\frac{d\sigma}{d\Omega}({\bf k}\beps\rightarrow{\bf k}'\beps')}.
\end{equation}  
Here, the light incidence is supposed to be perpendicular to the
surface $A$ of the medium which will be taken to become the half space
$z>0$ as $A\rightarrow \infty$.  The average differential cross
section is determined by Fermi's golden rule which reads
\begin{equation}
\mv{\frac{d\sigma}{d\Omega}({\bf k}\beps\rightarrow{\bf k}'\beps')}
= 
\frac{L^6\omega^2}{4\pi^2} 
\mv{|T({\bf k}'\beps',{\bf k}\beps)|^2}.
\label{crosssecdiff.eq} 
\end{equation}
The square of the transition operator means explicitly $|T({\bf k}' 
\beps',{\bf k}\beps)|^2=\bra{{\bf k}\beps}T(\omega+i0)^\dagger\ket{{\bf k}' 
\beps'}\bra{{\bf k}'\beps'}T(\omega+i0)\ket{{\bf k}\beps}$ 
and acts on
the atomic states only.  Note that the factor $L^6$
cancels with the inverse factor coming 
from the squared transition operator, so 
that the quantization volume finally disappears.

For single scattering,
eqs.~(\ref{singleamp.eq}) and (\ref{scattensor.eq}) 
show that the  internal average 
has to be taken over the square of the
(dimensionless)  scattering operator:
\begin{equation}
\mv{|\bbeps'\cdot{\bf \hat t}\cdot\beps|^2}_{\mathrm int}=
                {\mathrm Tr}[\rho
                (\bbeps\cdot {\bf d})(\beps' \cdot {\bf d})
                (\bbeps'\cdot {\bf d})(\beps \cdot {\bf d})]. 
\label{singleint.eq}
\end{equation}
It is crucial that the average be taken over the square of the
scattering tensor. This is not equivalent to taking the square of the
average which is essentially the polarizability 
(\ref{polarizability.eq}). Again, the
trace over a scalar density matrix will select the scalar part of the averaged
operator. But, as becomes evident in App.~\ref{ITOs.sec},  
now the antisymmetric and symmetric traceless parts can combine
with their counterpart in the direct product and contribute a non-trivial 
scalar component.      

In the double scattering situation, 
the two atoms are coupled by the intermediate
photon.  Let 
\begin{equation}
\hat t_{21}=\bbeps'\cdot {\bf \hat t}_2\cdot\Delta\cdot{\bf \hat t}_1\cdot
\beps
\label{t21op.eq}
\end{equation}
be a short-hand notation for the contracted double scattering operator for the
direct path, and $\hat t_{12}$ for the reverse path.  
The ladder contribution to the double
scattering intensity, just like in the classical case, is given by the
average sum of the squares of the two amplitudes, 
\begin{equation}
{\Tr}[\rho_{12} (|\hat t_{21}|^2 +|\hat t_{12}|^2)].
\label{doubleladderav.eq}
\end{equation}  
Here, $\rho_{12}$ is the two-scatterer density matrix.  The crossed
contribution is obtained, again in perfect analogy to the classical case, 
by the interference between the  
direct and reverse 
amplitude
\begin{equation}
 \Tr[\rho_{12}
(\hat t_{12}\bar{\hat t}_{21}\exp[i({\bf k}+{\bf k}')\cdot {\bf r}_{12}]+
(1\leftrightarrow2))].
\label{doublecrossedav.eq}
\end{equation} 
Since the atoms are uncorrelated, the density
matrix factorizes, $\rho_{12}=\rho_{1}\otimes\rho_{2}$. Furthermore, 
$\rho_\alpha=\rho$ since the atoms are identically distributed.  The
two-atom scattering operator 
(\ref{t21op.eq}) 
is not factorized in terms of elementary scalar products. 
But by expliciting the transverse projector
$\Delta_{ij} = \delta_{ij}- \hat n_i \hat n_j$, it becomes 
\begin{equation}
\hat t_{21}= (\bbeps'\cdot{\bf d}_2)[({\bf d}_2\cdot 
{\bf d}_1) -({\bf \hat n}\cdot{\bf d}_2)({\bf \hat n}\cdot{\bf
d}_1)](\beps\cdot {\bf d}_1).
\end{equation}  
All averages (\ref{singleint.eq}), (\ref{doubleladderav.eq}) and  
(\ref{doublecrossedav.eq}) 
can then be expressed as linear combinations of the one-atom trace 
\[{\mathrm Tr} [\rho ({\bf x}_4\cdot{\bf d}) 
({\bf x}_3\cdot{\bf d})({\bf x}_2\cdot{\bf d})({\bf x}_1\cdot{\bf d})]\]
where the fixed vectors ${\bf x}_\alpha$ stand for 
$\beps,\bbeps,\beps',\bbeps',{\bf\hat  n},$ or
${\bf d}'$ (the dipole operator of the other atom).
Rather than calculating each term separately, we determine 
the one-atom trace
for four arbitray ${\bf x}_\alpha$ and later substitute what is 
required by the single and double scattering terms.

\subsection{The single scattering vertex}

We proceed to calculate the dimensionless trace function 
\begin{equation} 
{\mathcal T}({\bf x}_\alpha) = \frac{1}{M_J}
{\Tr}
[\rho ({\bf x}_4\cdot{\bf d})({\bf x}_3\cdot{\bf d})({\bf x}_2\cdot{\bf d})
({\bf x}_1\cdot{\bf d})].
\end{equation}
It depends linearly on the components
of the ${\bf x}_\alpha$, albeit in a complicated manner, involving the
characteristics of the transition and the elements of the density
matrix. A systematic way of evaluating the trace is a development in
terms transforming under irreducible representations of the rotation
group~\cite{Blum81}.  We shall explicit the solution in the simplest
case when the atom is distributed with equal
probability over its internal substates. Since the corresponding 
density matrix is then proportional to unity and  
therefore a
scalar under rotations, the trace selects the scalar part of the
averaged operator. The result can only be a function of the scalar
products $({\bf x}_\alpha \cdot {\bf x}_\beta)$, of the most general form
\begin{equation}
\begin{split}
  {\mathcal T}({\bf x}_\alpha) & = w_1({\bf x}_1\cdot{\bf x}_2)
({\bf x}_3\cdot{\bf x}_4)  
  + w_2({\bf x}_1\cdot{\bf x}_3)({\bf x}_2\cdot{\bf x}_4) 
\\ & \quad 
 + w_3({\bf x}_1\cdot{\bf x}_4)({\bf x}_2\cdot{\bf x}_3).
\end{split}
\label{trace.eq} 
\end{equation}
The coefficients $w_i$ are calculated explicitly using the standard
techniques of irreducible tensor operators (details are given in
appendix~\ref{ITOs.sec}):
\begin{equation}
w_1  = \frac{s_0-s_2}{3}, \  
w_2  = \frac{s_2-s_1}{2}, \ 
w_3 = \frac{s_1+s_2}{2}
\label{deft.eq}
\end{equation}
where
\begin{equation}
s_K=3(2\Je+1)
\left\{\begin{array}{ccc}
        1&1&K\\
        J&J&\Je
\end{array}\right\}^2.
\label{defs.eq}
\end{equation}
The ``$6J$''-symbols~\cite{Edmonds60} 
(or Wigner coupling coefficients) contain all
essential informations about our problem. 
They are the simplest scalar
quantities that can be constructed from the basic ingredients $J$,
$\Je$, $1$ (the rank of the vector operator ${\bf d}$)
and $K$ (the tensor ranks of the irreducible components of the scattering
operator). The ``$6J$''-symbols introduce useful selection rules:
\begin{enumerate}
\item[$(i)$] $|J-\Je|\le 1$, the usual selection rule for a dipole
  transition;
\item[$(ii)$] $0\le K\le 2$: the scattering operator is the 
direct product of two
vector operators and thus has irreducible components of rank $K=0,1,2$. 
In other
words, the change of the atomic angular momentum 
is limited to $|m'-m|\le 2$ for the one photon scattering;
\item[$(iii)$] $K\le 2J$: the ground state degeneracy determines which tensor
rank comes into play. For $J=0$, $K=0$ and thus only
  Rayleigh transitions $m'-m=0$ are possible; for $J=1/2$, $K=0,1$ and
  degenerate Raman transitions with $|m'-m|=1$ become possible; for
  $J\ge1$, $K=0,1,2$ and all possible transitions $|m'-m|\le 2$ can
  take place.
\end{enumerate}
A sum rule over $K$ for the ``$6J$''-symbols 
implies that the $w$ coefficients are not
independent but obey
\begin{equation}
\label{eq:sum_rule}
w_1+w_2+3w_3 = 1
\end{equation}
for arbitrary $J,\Je$. 
Explicit formulae for the $w_i$ are contained in app.~\ref{wvalues.sec}.

We introduce a diagrammatical representation for the trace function
(\ref{trace.eq}):
  \begin{equation}
\vertex{{\bf x}_1}{{\bf x}_2}{{\bf x}_3}{{\bf x}_4} = 
     w_1 \horizontal{1\,}{\,2}{\,3}{4\,}
  +  w_2 \diagonal{1}{\,2}{\,3}{4}
  +  w_3 \vertical{1}{2}{3}{4}
\label{boxdef.eq}
\end{equation} 
This four-point intensity vertex is the weighted sum of the three
pairwise contractions between the vectors ${\bf x}_\alpha$.  A factor $w_1$
comes in for the horizontal pairwise contraction $({\bf x}_1 \cdot {\bf x}_2)
({\bf x}_3\cdot{\bf x}_4)$, a factor $w_2$ for a diagonal pairwise
contraction, and a factor $w_3$ for a vertical pairwise contraction.
It ressembles Wicks's theorem known from Gaussian 
integration~\cite{Wick50}, but here, the
weights of the possible contractions are not equal. As in quantum field
theory, this diagrammatic representation 
proves especially useful for the systematic description of higher-order
scattering (cf.~Sec.~\ref{double.sec}).

\subsection{The single scattering contribution}
\label{single.sec}

Using $\beps = \bar{\bf x}_1={\bf x}_4$ and 
$\beps' = {\bf x}_2=\bar{\bf x}_3$ in (\ref{trace.eq}), the internal
average 
(\ref{singleint.eq}) for the 
single scattering contribution becomes
\begin{equation}
\mv{|\bbeps'\cdot{\bf \hat t} \cdot \beps|^2}_{\mathrm int}
 =M_J \left(w_1|\bbeps'\cdot\beps|^2+w_2|\beps'\cdot\beps|^2+w_3\right)
\end{equation}
The average differential cross-section for single scattering on an
unpolarized atom is
\begin{equation}
\mv{\frac{d\sigma_{\mathrm S}}{d\Omega}}_{\!\!\mathrm int}
=\frac{3\sigma_{\mathrm tot}}{8\pi} 
\left(w_1|\bbeps'\cdot\beps|^2+w_2|\beps'\cdot\beps|^2+w_3\right)
\label{singlediffcross.eq}
\end{equation}
in terms of the total scattering cross-section (\ref{crossection.eq}).  
Using this expression, we see that the sum 
rule (\ref{eq:sum_rule}) simply represents flux conservation. 
All angular information is contained in the squared moduli of
polarization contractions. Since these expressions are even in the
scattering angle $\theta$, the mean value $\mv{\cos
  \theta}=\int d\theta \mv{d\sigma/d\Omega} \cos\theta$ vanishes, 
justifying that the transport mean free
path $\ell_{\mathrm{tr}}=\ell/(1-\mv{\cos\theta})$ is equal to the
scattering mean free path $\ell$.  

To determine the bistatic coefficient now means to average
eq.~(\ref{singlediffcross.eq}) over position.  We assume a
semi-infinite, homogenous medium of independently distributed atoms.
The single scattering bistatic coefficient then is
\begin{equation}
\gamma_{\mathrm S}=\frac{4\pi n}{A}\int_{z>0}   \!\!\!\!\!\!\! d^3{r} \;
  \mv{\frac{d\sigma_{\mathrm S}}{d\Omega}}_{\!\!\mathrm int} 
   e^{-2z/\ell}.
 \label{defbistatic.eq}
\end{equation}
The exponential
takes account of the extinction of incoming and scattered light with
the mean free path $\ell$ inside the scattering medium.  Since the
differential cross-section is independent of the position, the
integral is readily calculated and we find 
\begin{equation}
\gamma_{\mathrm S} =\frac{3}{4}\
 \vertex{\beps}{\bbeps'}{\beps'}{\bbeps} = 
\frac{3}{4}\left(w_1|\bbeps'\cdot\beps|^2 + w_2|\beps'\cdot\beps|^2 +
 w_3\right) .
\label{gamma_S.eq}      
\end{equation}
The coefficients $w_i(J,\Je)$ carry the weights of the different
contractions of the polarization vectors (for detailed expressions,
see app.~\ref{wvalues.sec}).  In the case of a transition $J=0$,
$\Je=1$, these coefficients are simply $(w_1,w_2,w_3)=(1,0,0)$. So one
recovers exactly the classical expression \cite{vanTiggelen90}
\begin{equation}
\gamma_{\mathrm S}=  \frac{3}{4}
\clvertex{\beps}{\bbeps'}{\beps'}{\bbeps} 
     =\frac{3}{4}|\bbeps'\cdot\beps|^2.
\end{equation} 
In the classical diagram, the upper line, read from left to right,
signifies scattering of the wave amplitude, and the lower line, read
from right to left, signifies scattering of the complex conjugate
amplitude by the same scatterer (identified by the dashed line). The
only possible connection is horizontal, giving the factor
$|\bbeps'\cdot\beps|^2$ that implies a dipole radiation pattern.  For
atoms, however, the coefficients $w_2$ and $w_3$ come into play and
lead to contributions in the $\lperp$ channel (where
$\bbeps'\cdot\beps=\beps'\cdot\beps=0$) as soon as $J\ge 1/2$.  When
$J\ge 1$, there is a signal even in the helicity preserving
backscattering channel $\hpar$ (where $\bbeps'\cdot\beps=0$,
$\beps'\cdot\beps=1$).  Now we see why the polarizability 
(\ref{polarizability.eq}) is not
sufficient to describe the scattering by a degenerate transition. The
polarizability is essentially the average scattering tensor, a
two-point vertex connecting the incoming to the outgoing
polarization. If the atom is distributed with equal probability over
its substates, the polarizability then is diagonal and defines a
purely horizontal contraction proportional to $|\bbeps'\cdot
\beps|^2$.  But the internal structure of an atom allows also for
diagonal and vertical connections in the single scattering diagram:
the classical line stretches to a two dimensional ribbon.

\subsection{The double scattering contributions} 

\label{double.sec}

Van Tiggelen {\em et al.}~\cite{vanTiggelen90} have calculated the
double-scattering contribution to the ladder bistatic
coefficient in the backward direction,
\begin{equation} 
\gamma_{\mathrm L2}=\frac{9}{16\pi A\ell^2}\int_{z_{1,2}>0}
\!\!\!\!\!\!\!\!\!  d^3{r}_1 d^3{r}_2 \;\; 
\frac{e^{-(z_1+r_{12}+z_2)/\ell}}{r_{12}^2}
\: P_{\mathrm L2}
\label{gammaL2.eq}
\end{equation} 
for classical point scatterers in a half-space, within the weak scattering
limit $k\ell \gg 1$ and in the far field approximation
$kr_{12} \gg 1$.  Here, the exponential describes the attenuation of
incident, intermediate and scattered light with 
mean free path $\ell$. 
For classical dipole point scatterers, the polarization kernel is given by
$P_{\mathrm L2}^{(\mathrm{cl})}=|\bbeps'\cdot\Delta\cdot\beps|^2$.  
For atomic scatterers
under the same conditions, eq.~(\ref{gammaL2.eq}) remains valid. As
all information about the internal structure is connected to the
polarization, only the polarization kernel has to be generalized. Keeping
track of all factors, it
follows from Sec.~\ref{ensembleav.sec} that the polarization kernel is
given as the internal average (\ref{doubleladderav.eq}) over the square of the
dimensionless double scattering operator:
\begin{equation}
P_{\mathrm L2}= M_J^{-2}\mv{
|\bbeps'\cdot {\bf \hat t}_2\cdot\Delta\cdot{\bf \hat  t}_1\cdot
\beps|^2}_{\mathrm int}.
\end{equation}
Using eqs.~(\ref{t21op.eq}) to (\ref{boxdef.eq}), 
it is represented by the generalized ladder diagram
\begin{equation}
P_{\mathrm L2}= \doubleladder{\beps}{\bbeps'}{\beps'}{\bbeps } .
\label{ladderbox.eq}
\end{equation}
This double-scattering ladder diagram is the product of two single
scattering diagrams 
(\ref{boxdef.eq}) connected by the polarization propagator
$\Delta_{ij}=\delta_{ij}-\hat n_i \hat n_j$, one for the amplitude
(upper line) and one for its complex conjugate (lower line). The
diagram is evaluated using the following rules. Each scattering box
yields three pairwise contractions: horizontal with weight $w_1$,
diagonal with weight $w_2$ and vertical with weight $w_3$.  Now choose
$w_i w_j$ for the two boxes and contract the vectors 
accordingly. For example, $w_1^2$ comes with the twofold horizontal
contraction $|\beps\cdot\Delta\cdot\bbeps'|^2$;  
$w_1 w_2$ and $w_2 w_1$ both
give $|\beps\cdot\Delta\cdot\beps'|^2$.  For the vertical connections
involving factors of $w_3$, one has to use that 
the polarization propagator is a projector,
$\Delta\cdot\Delta=\Delta$, and its total contraction (arising for
$w_3^2$) is $\sum_{i}\Delta_{ii}=2$. Finally
\begin{equation}
\begin{split}
P_{\mathrm L2}  & = 
	(w_1^2+w_2^2)|\bbeps'\cdot\Delta\cdot\beps|^2 + 2w_1 w_2
		|\beps'\cdot\Delta\cdot\beps|^2  \\
   & \quad + (w_1+w_2)w_3 
	[(\bbeps\cdot\Delta\cdot\beps) + (\bbeps'\cdot\Delta\cdot\beps')]
  + 2w_3^2
\end{split}
\label{PL2.eq}
\end{equation}
For classical dipole scatterers, modeled by a transition $J=0$,
$\Je=1$, one has $(w_1,w_2,w_3)=(1,0,0)$ and recovers the known result
\begin{equation}
P_{\mathrm L2}^{(\mathrm{cl})} 
= \cldoubleladder{\beps}{\bbeps'}{\beps'}{\bbeps }
=|\bbeps'\cdot\Delta\cdot\beps|^2  .
\label{cladderbox.eq}
\end{equation}

The crossed bistatic coefficient for the double-scattering contribution
as calculated by van Tiggelen {\em et al.}~\cite{vanTiggelen90} under
the same assumptions is
\begin{equation}
\begin{split}
  \gamma_{\mathrm C2}=\frac{9}{16\pi A\ell^2} \int_{z_{1,2}>0}
  \!\!\!\!\!\!\!\! d^3{r}_1 d^3{r}_2 & \;
  \frac{e^{-(s z_1+r_{12}+s z_2)/\ell}}{r_{12}^2} 
\\ &\times 
\cos[({\bf k}+{\bf k}')\cdot{\bf r}_{12}] \; P_{\mathrm C2}
\label{gammaC2.eq}
\end{split}
\end{equation} 
where $s=\frac{1}{2}(1+1/\cos\theta)$. For classical dipole point scatterers,
the crossed and ladder polarization kernels are equal,  
$P_{\mathrm C2}^{(\mathrm{cl})} = P_{\mathrm L2}^{(\mathrm{cl})} $.  
This assures that in the backscattering direction
($\theta=0$, ${\bf k}'=-{\bf k}$), crossed and ladder intensities are
equal (the strict equality for all polarizations is characteristic of
double scattering; for higher scattering orders, equality of crossed
and ladder is only given for parallel polarizations).  In the case of
atomic scatterers, eq.~(\ref{gammaC2.eq}) remains valid, but the
polarization kernel has to be generalized.  Casting the internal
average (\ref{doublecrossedav.eq}) in diagrammatical form,
it is
 \begin{equation}
P_{\mathrm C2}= \doublecrossed{\beps}{\bbeps'}{\beps'}{\bbeps} \quad .
\label{crosseddiagram.eq}
\end{equation}   
The crossed diagram is evaluated efficiently using the contraction
rules defined above for the ladder diagram. Explicitly,
\begin{equation}
\begin{split}
  P_{\mathrm C2} & = 
 (w_1^2+w_3^2)|\bbeps'\cdot\Delta\cdot\beps|^2
  + 2w_1 w_3 (\bbeps'\cdot\Delta\cdot\beps')(\beps\cdot\Delta\cdot\bbeps) \\
  & + (w_1+w_3)w_2 [(\beps\cdot\beps')(\bbeps\cdot\Delta\cdot\bbeps')+
  (\bbeps\cdot\bbeps')(\beps\cdot\Delta\cdot\beps')]\\
 & + 2w_2^2|\beps'\cdot\beps|^2 
\label{PC2.eq}  
\end{split}
\end{equation}
Obviously, the crossed kernel is not equal to the ladder kernel,
$P_{\mathrm C2}\ne P_{\mathrm L2}$. What is the relation between the
two? In the 
classical theory, one habitually uses time reversal invariance to
reduce one to the other: returning the lower line of the crossed
diagram for classical point scatterers
\begin{equation}
P_{\mathrm C2}^{(\mathrm{cl})} = 
\cldoublecrossed{\beps}{\bbeps'}{\beps'}{\bbeps} \quad ,
\label{ccrosseddiagram.eq}
\end{equation}   
the connecting lines are straightened out, and the crossed diagram
becomes equal to the ladder diagram (\ref{cladderbox.eq}) 
in the parallel polarization channels $\bbeps'=\beps$. This is the
signature of reciprocity and assure a perfect interference contrast
in the backscattering direction.
But returning the bottom line of the generalized crossed diagram
(\ref{crosseddiagram.eq}), we find
\begin{equation} 
P_{\mathrm C2}= 
\twisteddoublecrossed{\beps}{\bbeps'}{\bbeps}{\beps'}
\label{returneddiagram.eq}
\end{equation}    
which differs from the ladder diagram (\ref{ladderbox.eq}), even if we
put $\bbeps'=\beps$. What has happened?  The ribbon that has replaced
the classical line cannot unwind and blocks the diagram topologically.
It blocks because the diagonal and the vertical contraction are not
equivalent: $w_2 \ne w_3$. 
Only in the case $J=0$, $\Je=1$, we have $w_2=w_3=0$, and one recovers
the correspondence to the classical point dipole scatterers.
Eq.~(\ref{deft.eq}) shows that $w_2=w_3$ if and only if $s_1=0$.  The
coefficient $s_1$ stems from the antisymmetric part of the scattering
operator (cf. app.~\ref{ITOs.sec}).  As the analysis of the double
scattering amplitude in Sec.~\ref{doubleamp.sec} already showed, it is
the antisymmetric part of the atomic scattering operator that is
responsible for the reduction of the backscattering enhancement 
in the parallel channels. 

\section{Enhancement factors and peak analysis for any atomic transition}  
\label{results.sec}

The results of the previous section enable us to calculate
analytically the intensity of polarized light scattered at first and
second order by atoms which are positioned randomly in a half-space. 
The spatial integrals in eqs.~(\ref{gammaL2.eq}) and (\ref{gammaC2.eq})
are 
challenging  because of the half-space geometry, 
but can be performed fully analytically -- see 
section \ref{appendix:crossed}.  
We therefore obtain the various enhancement factors at backscattering
as well as the shape of the backscattered cone fully analytically.
In this section, we analyze the contributions of single and double
scattering to the backscattered intensity and determine the
second-order enhancement factor as a function of the polarization
channel and the atomic dipole transition.

\subsection{Single scattering background}

\begin{figure}
  \centering \includegraphics[width=0.45\textwidth]{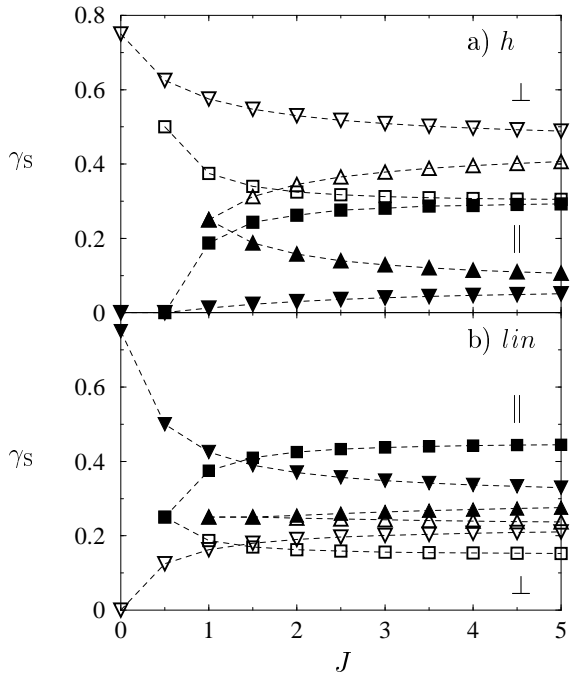}
\caption{Single scattering intensity in terms of the bistatic coefficient 
  (\ref{gamma_S.eq}) in the backscattering direction 
  as a function of the ground state angular momentum $J$: 
a) preserved and flipped helicity in the circular polarization channels, 
b) parallel and perpendicular polarization in the linear channels.  
 Full symbols: parallel channels ($\hpar$ and $\lpar$).  Open symbols:
  perpendicular channels ($\hperp$ and $\lperp$).  \transitiontypes.}
\label{single.fig} 
\end{figure}

The single scattering intensity in terms of the bistatic coefficient
$\gamma_{\mathrm S}$  
is given by eq.~(\ref{gamma_S.eq}) as a function of the ground state angular
momentum $J$, the transition type $\Je-J=0,\pm 1$ and the
polarization vectors.  Fig.~\ref{single.fig} shows $\gamma_{\mathrm S}$ for all
transition types and the four standard polarization configurations.
 
For classical point dipole scatterers, the single scattering bistatic
coefficient is $\gamma_{\mathrm S}=\frac{3}{4}$ in the channels $\hperp$ and
$\lpar$ (corresponding to the reflection from a mirror) and
$\gamma_{\mathrm S}=0$ in the channels $\hpar$ and $\lperp$ 
\cite{vanTiggelen90}.
Fig.~\ref{single.fig} reproduces these values for the transition
$J=0$, $\Je=1$.  As explained in Sec.~\ref{singleamp.sec}, for $J>0$
degenerate Raman transitions become possible and open the classically
forbidden channels: the first signal is obtained in the $\lperp$
channel for $J=\frac{1}{2}$ and in the $\hpar$ channel for $J=1$.

In all four polarization channels, the graphs of the two transition
types $\Je=J\pm1$ (upward and downward triangles) tend towards the
same value as $J \to \infty$. Indeed, as shown in
appendix~\ref{wvalues.sec}, the coefficients $w_i$ for these two
transition types have the same limit, corresponding to asymptotically
equal Clebsch-Gordan coefficients.

Two main conclusions are to be drawn from Fig.~\ref{single.fig}: 
\begin{itemize}
\item[$(i)$] A degeneracy of the atomic dipole transition leads to a
single scattering contribution to the backscattered intensity {\em in all
  four polarization channels} (with the only exception 
$J=\frac{1}{2}$ in $\hpar$); this background signal therefore
cannot be eliminated by polarization analysis and reduces the
observable height 
(\ref{efactor.eq}) 
of the coherent backscattering peak;  
\item[$(ii)$] The
intensity in the $\hpar$ and $\lperp$ channels always stays below the
intensity in the $\hperp$ and $\lpar$ channels, respectively; the
single scattering contribution thus is always minimized by choosing
the classically forbidden channels.
\end{itemize}

\subsection{Double scattering interference contrast}

\begin{figure}
  \centering \includegraphics[width=0.4\textwidth]{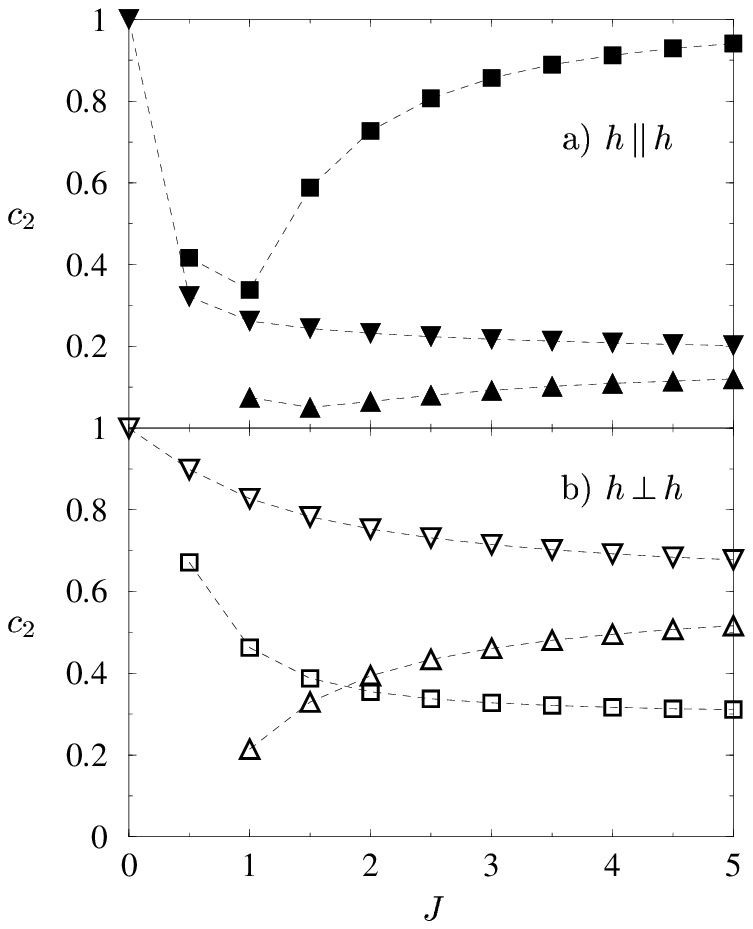}
\caption{Contrast of double backscattering interference
(\ref{contrast.eq}) as 
  function of the ground state angular momentum $J$ for circular
  polarizations: a) conserved helicity, b) flipped helicity.
  \transitiontypes.}
\label{doublehel.fig}
\end{figure}

\begin{figure}
  \centering \includegraphics[width=0.4\textwidth]{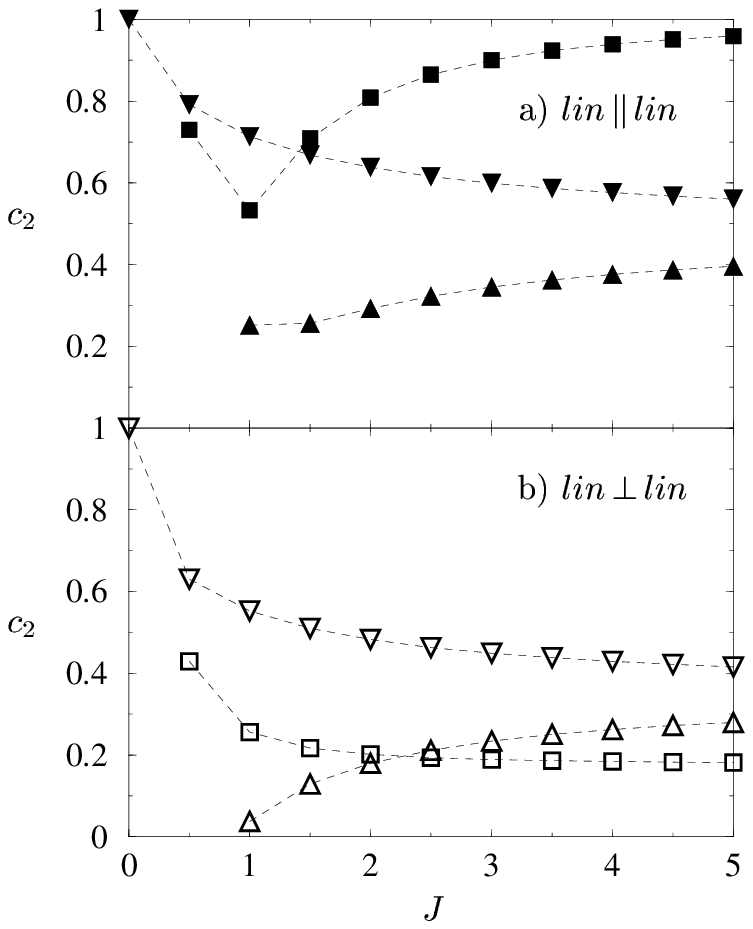}
\caption{Contrast of double backscattering interference
(\ref{contrast.eq}) as 
  function of the ground state angular momentum $J$ for linear
  polarizations: a) parallel polarizations, b) perpendicular
  polarizations.  \transitiontypes.}
\label{doublelin.fig}
\end{figure}

The contrast of second order backscattering interference,
\begin{equation}
c_2=\frac{\gamma_{\mathrm C2}(0)}{\gamma_{\mathrm L2}},
\label{contrast.eq}
\end{equation}
is determined by the crossed and the ladder bistatic coefficients,
given in eqs.~(\ref{gammaL2.eq}) and (\ref{gammaC2.eq}) as integrals
over the generalized polarization kernels (\ref{PL2.eq}) and
(\ref{PC2.eq}), respectively.  These integrals can be evaluated
analytically, and their expressions as functions of $J$ and $\Je$ in
the four polarization channels are contained in
appendix~\ref{height.sec}. 
Here, we plot  the interference contrast $c_2$ in
Figs.~\ref{doublehel.fig} and \ref{doublelin.fig} 
in the four standard polarization channels as a
function of the ground state angular momentum $J$.

Two features of Fig.~\ref{doublehel.fig} are particularly striking:
Firstly, a perfect contrast $c_2=1$ is obtained solely for the
transition $J=0$, $\Je=1$ corresponding to classical point dipole
scatterers.
The degeneracy of the atomic transition then degrades the
contrast considerably. For instance, in the channel of preserved
helicity ($\hpar$) and for a transition of type $\Je=J+1$ (full
downward triangles in Fig.~\ref{doublehel.fig}), the contrast drops
to about $0.3$ already at
$J=\frac{1}{2}$ and takes typical values of $0.2$. Secondly, the
channel $\hperp$ can offer a contrast up to three times higher than
the channel $\hpar$, depending on the transition type and the
degeneracy of the ground state.

Figs.~\ref{doublehel.fig}b) and \ref{doublelin.fig}b) show that in the
crossed channels ($\hperp$ and $\lperp$), the contrast is always
maximized for a transition type $\Je =J+1$. But in the parallel
channels ($\hpar$ and $\lpar$) and for larger values of $J$, the
contrast is optimized for $\Je=J$.  In the limit
$\Je=J\to\infty$, the contrast $c_2$ even approaches $1$. A contrast
of $1$ indicates that the antisymmetric part of the scattering tensor
vanishes.  Indeed, the Clebsch-Gordan coefficients 
display a symmetry that suppresses the antisymmetric
part of the scattering tensor as $\Je=J\rightarrow \infty$.

\subsection{Backscattering enhancement factor}

Fig.~\ref{single.fig} shows that the smallest single scattering signal
is obtained in the $\hpar$ channel for $\Je=J+1$. This
configuration could also be expected to render the best enhancement
factor.  However, Fig.~\ref{doublehel.fig} shows that the interference
contrast in this configuration is particularly low.  As will indeed be
seen in this section, the choice of $\Je=J+1$ and $\hpar$ does not
guarantee an optimized backscattering enhancement. Depending on the
degeneracy, the crossed channel or the transition type $\Je=J$
can offer a better interference contrast and lead to a higher
enhancement factor.

An enhancement factor up to second order,
\begin{equation}
\alpha=1+\frac{\gamma_{\mathrm C2}(0)}{\gamma_{\mathrm S}+\gamma_{\mathrm L2}},
\label{doubleenhance.eq}
\end{equation}
combines the single and double scattering
contributions. It has to be pointed out, however, that its exact value
may not be compared directly to experimental results. Indeed, either
the scattering medium has the semi-infinite geometry of a half-space,
but then third and higher scattering orders cannot be neglected. Or it
has the finite geometry of a laser-cooled atomic cloud which truncates
higher scattering orders, but then the relative weight between single
and double scattering is modified. Numerical simulations can determine
the role of restricted geometry and are currently under study.
Preliminary results show that the ratio of the double-scattering
crossed intensity
to the ladder intensity is almost independent of the shape
of the medium. In other words, it is the 
internal
structure 
which is essential for the low contrast of
the interferences, not the spatial arrangement of the various atoms.
Thus, the present analytical calculation permits to follow how
the effects of single scattering background and reduction of
interference contrast combine to result in small enhancement factors.

\begin{figure}
  \centering \includegraphics[width=0.4\textwidth]{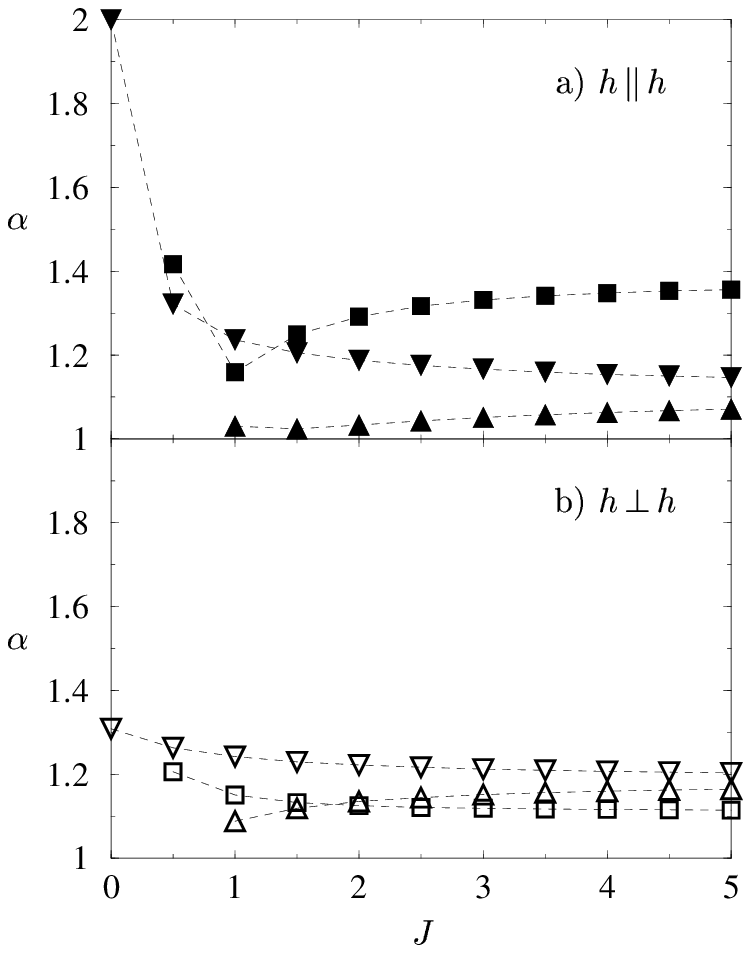}
\caption{The second-order backscattering enhancement factor 
(\ref{doubleenhance.eq}) 
  as function of the ground state angular momentum $J$ for
  circular polarizations: a) conserved helicity, b) flipped helicity.
  \transitiontypes.}
\label{enhancehel.fig}
\end{figure}

\begin{figure}
  \centering \includegraphics[width=0.4\textwidth]{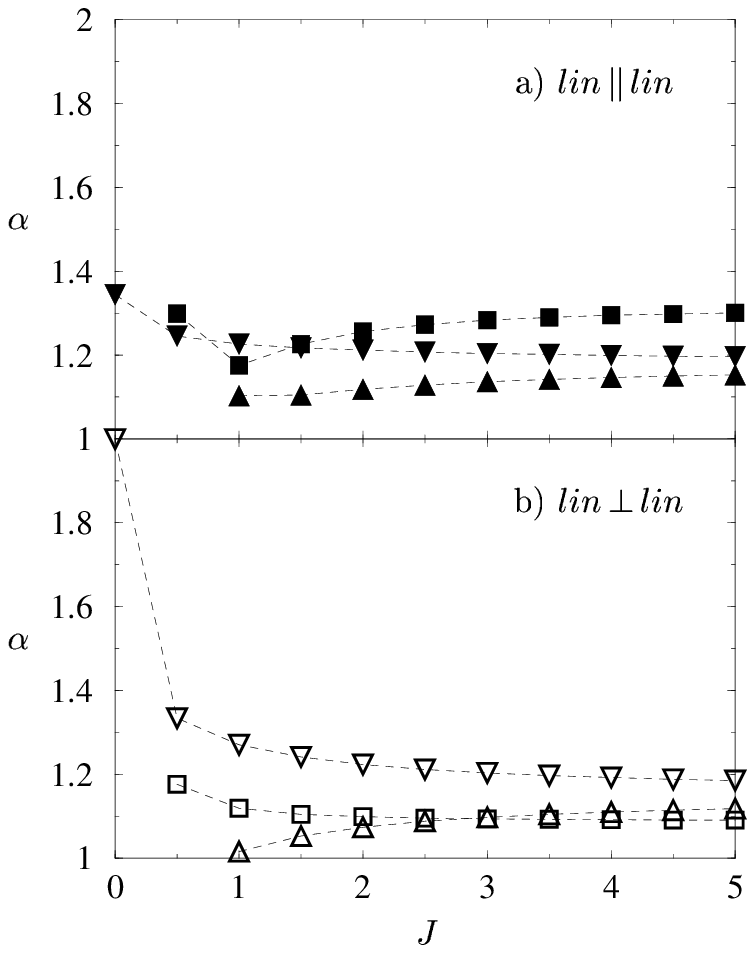}
\caption{The second-order backscattering enhancement factor 
  (\ref{doubleenhance.eq}) as 
   function of the ground state angular momentum $J$ for
  linear polarizations: a) parallel polarizations, b) perpendicular
  polarizations. \transitiontypes.}
\label{enhancelin.fig}
\end{figure}

Figs.~\ref{enhancehel.fig} and \ref{enhancelin.fig} show the
enhancement factor 
(\ref{doubleenhance.eq}) 
as a function of the ground state angular momentum
for the four standard polarization channels.

The difference between Figs.~\ref{doublehel.fig} and
\ref{enhancehel.fig} is given by the single scattering contribution,
shown in Fig.~\ref{single.fig}a).  In the channel of conserved
helicity ($\hpar$), the lowest single scattering intensity is observed
for the $\Je=J+1$ transition type, so that the enhancement factor,
following closely the interference contrast, drops from its classical
value $2$ to about $1.2$. The already poor contrast for the 
$\Je = J-1$ transition type is further reduced by single scattering. The
increasingly good contrast for $\Je=J$ at higher values of $J$
is counterbalanced by an important single scattering contribution, so
that the effective enhancement stays below $1.4$. We thus find that
the classical enhancement factor of $2$ in the helicity preserving
channel is irrevocably lost for atomic scatterers as soon as $J>0$.

Although the contrast of interference tends to be higher in the
$\hperp$ channel than in the $\hpar$ channel
(Fig.~\ref{doublehel.fig}), the single scattering contribution
(Fig.~\ref{single.fig}b)) also is more important, resulting in low
enhancement factors below $1.31$.
  
The enhancement factors in the linear channels, displayed in
Fig~\ref{enhancelin.fig}, show the same characteristics. With the only
exception of the $J=0,\Je=1$ transition in the 
$\lperp$ channel, we
find that all possible atomic transitions yield enhancement factors
below $1.35$.

The interplay between single scattering background and interference
contrast makes it difficult to predict in which configuration the
optimal enhanced backscattering can be measured.  Intuition formed
with classical scatterers would recommend a transition of type 
$\Je=J+1$ and the $\hpar$ channel. But for a high enough degeneracy of the
atomic transition, classical intuition turns out to be a bad
counsellor.  For $J=3$, $\Je=4$, the calculated effective enhancement
factor is higher in the $\hperp$ channel ($\alpha= 1.21$) than in the
$\hpar$ channel ($\alpha=1.17$). This had first been observed
experimentally~\cite{Labeyrie99} and remained puzzling until taking
into account the atomic internal structure~\cite{Jonckheere00}.

Fig.~\ref{enhancehel.fig} indicates that an optimized enhancement is
expected for a transition $\Je=J$ in the $\hpar$ channel. However, a
direct experimental verification seems delicate because a transition
of type $\Je=J$ is not closed in general 
(the emission of a
photon from the excited level $\Je$ to a final level $\Je=J-1$ is
allowed).  These events cut off elastic
scattering paths and yield a high background intensity, unfavourable
for experimental detection.  An interesting exception to this rule is
the closed transition $J=\Je=\frac{1}{2}$, which has the additional
advantage that no single scattering background pollutes the $\hpar$
channel.
  
\subsection{Enhanced backscattering peaks for $J=3$, $\Je=4$}

\begin{figure}
  \centering \includegraphics[width=0.5\textwidth]{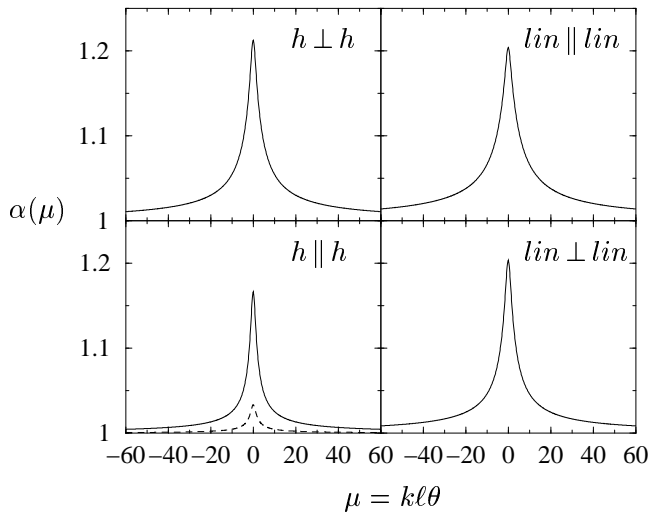}
\caption{The intensity enhancement 
  (\ref{alpha2.eq}) as a function of the reduced scattering angle
  $\mu=k\ell\theta$ for $J=3$, $\Je=4$ in the four polarization
  channels. In the linear channels, the intensity is scanned in the
  direction parallel to the incident polarization. In the helicity
  channels, the intensity is independent of the scan direction. 
 The dashed curve in the $\hpar$ channel shows CBS contribution from
Rayleigh transitions ($m'=m$) only. 
In this channel, the dominant contribution to the
CBS peak comes from Raman transitions between different substates
$m'\ne m$, see also Sec.~\ref{Raman.sec}.}  
\label{cone34.fig}
\end{figure}

The scattered intensity enhancement
\begin{equation}
\alpha(\mu) = 1 + 
 \frac{\gamma_{\mathrm C2}(\mu)}{\gamma_{\mathrm S}+\gamma_{\mathrm L2}}
\label{alpha2.eq}
\end{equation}
as a function of the reduced scattering angle $\mu=k\ell\theta$ for
any atomic transition and any polarization is given analytically in
terms of the bistatic coefficients (see app.~\ref{height.sec} for
details).  Fig.~\ref{cone34.fig} displays the backscattering peak
$\alpha(\mu)$ for the case that has been experimentally studied: 
the optical transition between two hyperfine levels  
($F=3$ and $F_{\mathrm e}=4$) of laser cooled Rubidium
atoms~\cite{Labeyrie99}. Hyperfine levels are characterized by a total 
angular momentum ${F}$ including the coupling with the nuclear spin;    
our analysis applies to any total angular momentum which we continue to note 
here by $J$. The highest peak arises in the channel of
flipped helicity ($\hperp$), the linear peaks are almost equivalent,
and the smallest peak is given for preserved helicity ($\hpar$). The
calculated enhancement factor in all four channels is of the order of
$1.2$. The experimentally measured enhancement factors are smaller
than the present values because the atomic cloud has neither a uniform
density nor the geometry of a half-space. Nevertheless, we stress that
the calculated peaks reproduce semi-quantitatively the experimental
ones as shown in~\cite{Jonckheere00}.

The shape of the CBS cone for atoms is 
similar to the
one for point dipole scatterers: the angular width is of the order of
$1/k\ell$ and can vary by a factor 2 depending on the polarization
channel. In the helicity channels, the CBS cone
is isotropic. In the linear channels, its presents an anisotropy which is
characteristic of polarization memory in low-order scattering
\cite{vanAlbada87}, see also App.~\ref{wings.sec}.  
In the $\lpar$ channel, it extends further in the direction
of the polarization than perpendicularly, reflecting the 
anisotropy of the dipole scattering cross-section in the Fourier plane.    

A small scattering angle is associated to endpoints of scattering
paths lying far apart. Conversely, short scattering paths dominate for
larger scattering angles.  Analytical results for double scattering
thus provide information about the wings of the backscattering peak
that are in principle measurable experimentally.  App.~\ref{wings.sec} 
contains the analytical expressions for the wings of
the backscattering peak in the four usual polarization channels. It
can be seen that the
intensity decreases as
$(k\ell\theta)^{-1}$ in all four channels, with coefficients depending
on $J$, $\Je$. Furthermore, the anisotropy in the linear polarization
channels, i.e., the dependence of the scattered intensity on the angle
between incident polarization and the direction of the scan, decreases
as the degeneracy of the atomic transition increases.  This reduction
of the anisotropy is consistent with the intuitive picture that
degenerate atomic transitions depolarize the incident light more
efficiently than dipole point scatterers and that the memory of direction as
seen in the backscattering anisotropy is lost more rapidly.

\section{Conclusions and outlook}

The internal structure of an atom determines its light scattering
properties. We have shown that a degeneracy of 
the atomic dipole transition
reduces the observable backscattering enhancement factor in two ways:
single scattering is present in all polarization channels and the
non-scalar part of the scattering tensor 
reduces the contrast of CBS interference. 
A complete analytical solution for the case
of unpolarized atoms has been presented together with a generalization
of the classical ladder and crossed diagrams to the case of atoms. An
immediate extension, under current study, is the application of the
present method to higher orders of scattering.  Finally, going beyond
the weak localization regime, further research is needed in order to
decide whether the the internal structure is not a substantial
difficulty in the quest for the strong localization of light in cold
atomic gases.

\section*{Acknowledgements}
We thank Robin Kaiser for fruitful discussions and Bart van Tiggelen 
for his constant interest.
We wish to thank Serge Reynaud for his useful advice in the early
stage of the work.  We also thank the PRIMA Research Group.
Laboratoire Kastler Brossel is laboratoire 
de l'Universit{\'e} Pierre et Marie
Curie et de l'Ecole Normale Sup{\'e}rieure, unit{\'e} mixte de
recherche 8552 du CNRS.

\onecolumn

\appendix
\section{Trace evaluation using irreducible tensor operators}
\label{ITOs.sec}

In the following, we employ the standard theory of irreducible tensor
operators as exposed in the textbooks by Edmonds~\cite{Edmonds60} and
Blum~\cite{Blum81}. We have to calculate
\begin{equation}
{\mathcal T}({\bf x}_\alpha)= \frac{1}{M_J}{\mathrm Tr}
[\rho_J ({\bf x}_4\cdot {\bf d})({\bf x}_3\cdot {\bf d})({\bf
x}_2\cdot {\bf d})({\bf x}_1\cdot {\bf d})]
\label{apptrace.eq}
\end{equation}
where ${\bf d}={\bf d}^{(1)}$ is the reduced dipole operator 
${\bf d}= {\bf D}/D$, an irreducible
tensor operator of rank $1$ acting upon the eigenstates $\ket{Jm}$ of
the angular momentum operators $J^2$ and $J_z$. 
Its reduced matrix element is by definition 
$\bra{\Je}|{\bf d}|\ket{J}= \sqrt{2\Je+1}$.  
We introduce the ratio of multiplicities $M_J=(2\Je+1)/3(2J+1)$ for
convenience.  The ${\bf x}_\alpha={\bf x}_\alpha^{(1)}$ are 
fixed free vectors commuting with $J^2,J_z,{\bf d}$. Therefore, the 
${\bf x}_\alpha$ are irreducible tensors of rank $1$, but not
operators, and the trace (\ref{apptrace.eq}) acts only on ${\bf d}$.
   
Let $O$ be an operator decomposed into its irreducible components,
\begin{equation}
O=\sum_{L,q}a_{Lq}O^{(L)}_q. 
\end{equation}
Its average $\mv{O}={\mathrm Tr} \rho O$ in a system described by a
density matrix $\rho$ can again be decomposed,
\begin{equation}
{\mathrm Tr} \rho O=\sum_{L,q}\frac{a_{Lq}}{\sqrt{2L+1}}
        \sum_{J,J'}\rho^{L}_q(J,J')
        \bra{J'}|O^{(L)}|\ket{J}. 
\label{tracedev.eq}
\end{equation} 
All angular information has been concentrated into the
coefficients $a_{Lq}$ and the components
\begin{equation}
\rho^{L}_q(J,J')=\sum_{m,m'}(-)^{L-J'-m}\langle JJ'{-m}m'|Lq \rangle
        \bra{Jm}\rho\ket{J'm'}
\end{equation} 
of the so-called statistical tensor operator.

If the system is distributed with equal probability over all substates
$\ket{Jm}$ for a given $J$, the density matrix with elements
$\bra{Jm}\rho\ket{J'm'}=(2J+1)^{-1}\delta_{JJ'}\delta_{mm'}$ is purely
scalar, and its only non-zero irreducible component is
$\rho^{0}_0=(2J+1)^{-1/2}$. The trace (\ref{tracedev.eq}) then reduces
to
\begin{equation}
{\mathrm Tr}\rho O= \frac{a_{00}}{\sqrt{2J+1}} \bra{J}|O^{(0)}|\ket{J}.
\label{trace0.eq}
\end{equation}

All we have to do now is to decompose the operator $O=({\bf x}_4\cdot
{\bf d})({\bf x}_3\cdot {\bf d})({\bf x}_2 \cdot {\bf d})
({\bf x}_1\cdot {\bf d})$  
into its irreducible components, determine the coefficient $a_{00}$
and the reduced matrix element $\bra{J}|O^{(0)}|\ket{J}$.  We begin
with decomposing the operator of second order 
$o_{21}=({\bf x}_2\cdot {\bf d})({\bf x}_1\cdot {\bf d})$. 
The scalar products can be expressed in any
basis, in the Cartesian basis as well as in irreducible components,
\begin{equation}
o_{21} = \sum_{K=0}^2\sum_{m=-K}^K (-)^{K-m}[x_2x_1]^{(K)}_{-m}
 [dd]^{(K)}_m.
\end{equation}
Here, $[A^{(k)} B^{(k')}]$ denotes the direct
product of two irreducible tensors.  The irreducible components of the
product are composed from the irreducible components of the factors,
\begin{equation}
[A^{(k)}B^{(k')}]^{(K)}_m=\sum_{r,s}\mv{kk'rs|Km} A^{(k)}_r B^{(k')}_s,
\end{equation}
using the Clebsch-Gordan coefficients $\langle kk'rs |
Km\rangle$.  Application of the inverse formula
\begin{equation}
A^{(K)}_m B^{(K')}_{m'}=\sum_{L,q}\mv{KK'mm'|Lq}
[A^{(K)}B^{(K')}]^{(L)}_q 
\end{equation}
to the product $O=o_{43} o_{21}$ leads to the
decomposition
\begin{equation}
O=\sum_{K,K',L,q}(-)^{K+K'-q}\left[[x_4x_3]^{(K)}[x_2x_1]^{(K')}\right]^{(L)}_{-q}\left[[dd]^{(K)}[dd]^{(K')}\right]^{(L)}_q
\end{equation}
which is a linear combination (sum over $K,K'$) of 
totally decomposed operators (sum over $L,q$).  
Under the trace according to eq.~(\ref{trace0.eq}), only
$L=q=0$ survives so that we are left with a sum of three terms
$K=K'=0,1,2$.

The reduced matrix element $ \bra{J}| O^{(0)}(K) |\ket{J} = \bra{J}|
[[dd]^{(K)}[dd]^{(K)}]^{(0)} |\ket{J}$ can be calculated using the
general formula
\begin{equation}
\bra{J'}|[A^{(k)}B^{(k')}]^{(k'')}|\ket{J}=
(-)^{k''+J+J'}(2k''+1)^{1/2}
  \sum_{J''}
        \left\{\begin{array}{ccc}
        k&k'&k''\\
        J&J'&J''
\end{array}\right\}
\bra{J'}|A^{(k)}|\ket{J''}
\bra{J''}|B^{(k')}|\ket{J}
\end{equation}
for the reduced matrix element of the direct product of two
irreducible tensor operators acting on the same
system~\cite{Edmonds60}.  Two iterated applications of this formula
yield
\begin{equation}\bra{J}|O^{(0)}(K)|\ket{J}
=(2\Je+1)^2 (-)^K
\left(\frac{2K+1}{2J+1}\right)^{1/2}
\left\{\begin{array}{ccc}
        1&1&K\\
        J&J&\Je
\end{array}\right\}^2.
\end{equation}

The last thing to do now is to evaluate 
\begin{equation}
  (-)^K(2K+1)^{1/2}a_{00}(K)=[x_1 x_2]^{(K)}\cdot[x_3 x_4]^{(K)}
\end{equation}
This scalar product of two irreducible tensors of rank $2$ can again
be written in any basis, in irreducible components as well as in
Cartesian components,
\begin{equation}
[x_1 x_2]^{(K)}\cdot[x_3 x_4]^{(K)}=\sum_{i,j}[x_1 x_2]^{(K)}_{ij}[x_3
x_4]^{(K)}_{ij}. 
\end{equation}
The cartesian components $[x_\alpha x_\beta]^{(K)}_{ij}$ are given
by the usual decomposition of matrices: for $K=0$, the scalar part or
trace
\begin{equation}
[x_\alpha x_\beta]^{(0)}_{ij}=\frac{1}{3} ({\bf x}_\alpha\cdot {\bf x}_\beta)
\delta_{ij},
\end{equation}
for $K=1$ the antisymmetric part
\begin{equation} 
[x_\alpha x_\beta]^{(1)}_{ij}=\frac{1}{2}(x_{\alpha i}x_{\beta
j}-x_{\alpha j}x_{\beta i})
\end{equation} 
and for $K=2$ the traceless symmetric part
\begin{equation} 
[x_\alpha x_\beta]^{(2)}_{ij}=\frac{1}{2}(x_{\alpha i}x_{\beta
j}+x_{\alpha j}x_{\beta i}) -
\frac{1}{3} ({\bf x}_\alpha \cdot {\bf x}_\beta) \delta_{ij}. 
\end{equation}
Putting everything together, we summarize
\begin{align}
  {\mathcal T}({\bf x}_i)&= 3(2\Je+1)\sum_{K} \left\{\begin{array}{ccc}
      1&1&K\\
      J&J&\Je
\end{array}\right\}^2
{\mathcal T}_K({\bf x}_i)
\nonumber\\
{\mathcal T}_0&=\frac{1}{3} ({\bf x}_1\cdot {\bf x}_2)({\bf x}_3\cdot {\bf x}_4)\\
{\mathcal T}_1&=\frac{1}{2} [({\bf x}_1 \cdot {\bf x}_4 )(
{\bf x}_2\cdot {\bf x}_3) -({\bf x}_1\cdot {\bf x}_3)({\bf x}_2\cdot {\bf x}_4)]\nonumber\\
{\mathcal T}_2&=\frac{1}{2} [({\bf x}_1 \cdot {\bf x}_4 )({\bf x}_2\cdot \bs
x_3)+({\bf x}_1\cdot {\bf x}_3)({\bf x}_2\cdot {\bf x}_4)] - {\mathcal
  T}_0.\nonumber
\end{align}
This form shows nicely that the scalar, antisymmetric and
traceless symmetric parts of the single scattering operator combine
with their counterparts in the direct product 
and contribute to the scalar trace.  
Regrouping of the different contractions leads to the vertex form 
(\ref{trace.eq}) presented
in Sec.~\ref{ensembleav.sec}.

\twocolumn
\section{Analytical expressions of the double scattering contributions} 
\label{height.sec}

\subsection{Values of transition-dependent coefficients}
\label{wvalues.sec}

For $\Je=J+1$, the transition-dependent coefficients 
(\ref{deft.eq}) are explicitly
\begin{equation}
w_i=\frac{1}{10(J+1)(2J+1)} \times 
        \begin{cases}
          6J^2+17J+10, & i=1\\
          -4J(J+2)   , & i=2\\
          J(6J+7) , & i=3
        \end{cases}
\end{equation}
For $\Je=J$,
\begin{equation}
w_i=\frac{1}{10J(J+1)} \times 
        \begin{cases}
          2J^2+2J+1,  &i=1\\
          2(J+2)(J-1),&i=2\\
          2J^2+2J+1, &i=3
        \end{cases}
\end{equation}
For $\Je=J-1$,
\begin{equation}
w_i=\frac{1}{10J(2J+1)} \times 
        \begin{cases}
          (6J+1)(J-1), & i=1\\
          -4(J+1)(J-1),& i=2\\
          (J+1)(6J-1), & i=3
        \end{cases}
\end{equation}
As pointed out in Sec.~\ref{double.sec}, the antisymmetric part of the
scattering tensor plays no role when $w_2=w_3$.  The only finite
values of $J$, $\Je$ for which this condition is fulfilled are
$J=0,\Je=1$, the case of the classical dipole point scatterer, where
$w_2=w_3=0$.  The coefficients take non-trivial values in the limit
$J\rightarrow\infty$:
\begin{equation}
(w_1,w_2,w_3)=\frac{1}{10} \times 
    \begin{cases}
      (3,-2,3) &, \Je=J\pm1 \\
      (2,2,2) &, \Je=J
        \end{cases}
\end{equation}
and we see that a non-trivial realization $w_2=w_3=\frac{1}{5}$ of a
vanishing antisymmetric part of the scattering tensor is given
asymptotically in the case $\Je=J\rightarrow\infty$.

\subsection{Ladder contribution}

The six-dimensional integral (\ref{gammaL2.eq}) with the generalized ladder
polarization kernel (\ref{PL2.eq}) can be exactly calculated.
The first (trivial) step is to use the translational invariance perpendicularly
to the incoming direction, and reduce it to an integral over the
three components of the interparticle vector ${\bf r}_{12}$ 
and over $(z_1+z_2)/2.$
In a second step, we use spherical coordinates
$(r_{12},\vartheta,\varphi)$ for 
 ${\bf r}_{12}$, where $\vartheta$ is the angle between the
$z$-direction and ${\bf r}_{12}$, and $\varphi$  the azimuthal angle
(in the circularly polarized case, the ladder
kernel is independent on $\varphi$). 
The integrals over $r_{12}$ and $(z_1+z_2)/2$ are then easily
performed, leading to the double scattering
ladder contribution:
\begin{equation}
\gamma_{\mathrm L2}=\frac{9}{32\pi}
\iint{\frac{\sin \vartheta \ P_{\mathrm L2}
(\vartheta,\varphi)\ d\vartheta d\varphi }{1+|\cos \vartheta|}}
\end{equation}
expressed as an integral over the direction 
${\hat n}=(\sin \vartheta \cos \varphi,\sin \theta \sin \varphi,
\cos \vartheta)$ 
of the interparticle vector. 
The kernel $P_{\mathrm L2},$ given by eq.~(\ref{PL2.eq}),
involves only simple trigonometric functions of $\vartheta$ and $\varphi,$
which makes the calculation of the integral easy. The result depends
of course on the incoming and outgoing polarizations $\beps$ and
$\beps'.$

We finally obtain: 
\begin{equation}
\gamma_{\mathrm L2}=\frac{9}{8}\left(l_1(w_1+w_2)^2 + l_2 w_1 w_2 +
l_3(w_1+w_2)w_3 + l_4 w_3^2 \right) .
\end{equation}     
Here, the terms $w_1^2+w_2^2$ have been completed to
$(w_1+w_2)^2$, simplifying all following expressions.  
The coefficients $w_i(J,\Je)$ carry the dependence on the atomic
transition, and the coefficients $l_i$ are given as functions of the
polarization channels: \renewcommand{\arraystretch}{1.3}
\begin{equation}
\begin{tabular}{|c|c|c|c|c|}
\hline
    &$\hpar$            &  $\hperp$     &$\lpar$        &$\lperp$       \\
\hline
$l_1$ &$\frac{5}{48}$   &$\ln 2-\frac{19}{48}$&$\ln 2-\frac{11}{32}$ &
$\frac{5}{96}$\\
$l_2$ &$2\ln 2-1$&$-(2\ln 2-1)$ &       $0$&$0$ \\
\hline
\end{tabular}
\label{laddercoeff12.eq}
\end{equation}
\renewcommand{\arraystretch}{1.0}
and 
\begin{equation}
l_3 = 2\ln 2-\frac{1}{2}, \qquad l_4 = 2\ln 2
\label{laddercoeff34.eq}
\end{equation}
in all four channels. The coefficient $l_1$ for the four
channels had been derived in ref.~\cite{vanTiggelen90}, the others
describe the generalization to the case of degenerate atomic
transitions.

\subsection{Crossed contribution} 

\label{appendix:crossed}
The calculation of the crossed contribution given by 
eqs.~(\ref{gammaC2.eq}) and
(\ref{PC2.eq}) follows the same lines. There is however 
a complication due to the $\cos[({\bf k}+{\bf k}')\cdot{\bf r}_{12}]$ 
term. We choose the spherical coordinates $(r_{12},\vartheta,\varphi)$
such that the $x$ axis is along $({\bf k}+{\bf k}')$, that is
in the direction of observation. The integral over the transverse
components of ${\bf r}_{12}$ and over $(z_1+z_2)/2$ and
$r_{12}$ yields the following result:
\begin{equation}
\gamma_{\mathrm C2}(\mu)=\frac{9}{32\pi}
\iint{\frac{\sin \vartheta\ P_{\mathrm C2}(\vartheta,\varphi)\ d\vartheta d\varphi}
{1+|\cos \vartheta|+\mu^2(1-|\cos \vartheta|)\cos^2\varphi}}
\end{equation}
where
\begin{equation}
\mu=\theta k \ell
\end{equation}
is the reduced scattering angle.

The kernel $P_{\mathrm C2}$ is a combination of simple
trigonometric functions of $\vartheta$ and $\varphi$. This makes
it possible to calculate easily the integral over $\varphi$, leading
to: 
\begin{equation}
  \gamma_{\mathrm C2}(\mu)=\frac{9}{8}\int_0^1 \!\!\!\! dx \; \frac{C(x;J,\Je)} 
{\sqrt{(1+x)^2+\mu^2(1-x^2)}}
\label{gammaCint.eq}
\end{equation}
where the crossed kernel $C(x;J,\Je)$ depends on the atomic transition
$J \rightarrow \Je$ via the coefficients $w_i$,
\begin{equation}
\begin{split}
C(x;J,\Je) & = (w_1+w_3)^2 c_1(x)+ w_1w_3  c_2(x)
\\ &\quad 
+  (w_1+w_3)w_2 c_3(x) +  w_2^2 c_4(x)
\end{split}
\end{equation}
and the functions $c_i(x)$ depend on the polarization channel:
\begin{equation}
  \renewcommand{\arraystretch}{1.3}
\begin{tabular}{|c|c|c|c|c|}
\hline
    &$\hpar$            &  $\hperp$     &$\lpar$        &$\lperp$       \\
\hline
$c_1(x)$ &$\frac{1}{4}(1-x^2)^2$&$\frac{1}{4}(1+x^2)^2$ &
    $\frac{1}{4}(1+x^2)^2+A_\parallel$ & $A_\perp$\\
$c_2(x)$ &$2x^2$                &$0$            &       $0$&$2x^2$      \\
$c_3(x)$ &$1+x^2               $&$0$    &       $1+x^2+B_\parallel$&$0$ \\
$c_4(x)$ &$2$           &$0$            &       $2$&$0$ \\
\hline
\end{tabular} 
\renewcommand{\arraystretch}{1.0}
\label{cofxtable}
\end{equation}
In the $\hperp$ channel, the only non-zero coefficient is $c_1(x)$. This
means that apart from a multiplicative factor $(w_1+w_3)^2$, the
backscattering peak for any atomic transition has exactly the same
shape as the classical peak. 
This is due 
to the fact that only Rayleigh transitions
contribute to the CBS peak in the $\hperp$ channel and that
the radiation diagram of such transitions is -- averaged over
the magnetic quantum number -- identical to the one
of classical point dipole scatterers. In all other channels, the form of the
backscattering peak itself is changed, be it only in minor ways.  In
the linear channels, a supplementary complication arises because the
intensity depends on the angle $\phi$ between the incident
polarization vector and the direction of the observation. $\phi=0$
corresponds to a scan parallel to the incident polarization vector,
$\phi=\frac{\pi}{2}$ to a scan perpendicular to the incident
polarization vector (in ref.~\cite{vanTiggelen90}, the opposite
convention is chosen). This anisotropy of the backscattering
enhancement, observed already for classical point scatterers, is
contained in the expressions 
\begin{align}
  A_\parallel & = \frac{(1-x^2)^2}{8}
  (1+X^2\cos{4\phi}) + \frac{1-x^4}{2} X \cos{2\phi}\nonumber\\
  B_\parallel & = (1-x^2)X
  \cos{2\phi}     \label{anisotropies.eq}\\
  A_\perp & = \frac{(1-x^2)^2}{8}(1-X^2\cos{4\phi}) \nonumber
\end{align}
with
\begin{equation}
X=1-2\frac{\sqrt{(1+x)^2+\mu^2(1-x^2)}-1-x}{(1-x)\mu^2}.
\end{equation}

Finally, the integral (\ref{gammaCint.eq}) can be calculated
analytically.
The expressions are rather complicated and we give them for
completeness. We obtain:
\begin{equation}
\begin{split} 
\gamma_{\mathrm C2}(\mu)  & = (w_1+w_3)^2 \gamma_1(\mu) +
w_1 w_3 \gamma_2(\mu) 
\\ & \quad  
   + (w_1+w_3)w_2 \gamma_3(\mu) + w_2^2 \gamma_4(\mu)
\end{split}
\end{equation}
where the
non-zero functions $\gamma_i(\mu)$ are given by the following
expressions: 

$\bullet$ In the $\hpar$ channel:
\begin{equation}
\begin{split}
\gamma_1(\mu) & = \frac{3}{256(1-\mu^2)^4}
	\big[ 32-176\mu^2-84\mu^4+18\mu^6 \\
 	& \quad +(-22+144\mu^2-17\mu^4)\sqrt{1+\mu^2} \\
 	& \quad +3\mu^4(48-16\mu^2+3\mu^4)F(\mu)\big] \\
\gamma_2(\mu) & =  \frac{9}{8(1-\mu^2)^2}
	\big[ -4-2\mu^2 + 3\sqrt{1+\mu^2} \\
	& \quad  + (2+\mu^4)F(\mu) \big]  \\
\gamma_3(\mu) & = \frac{9}{16(1-\mu^2)^2}
	\big[ -4-2\mu^2 + 3\sqrt{1+\mu^2} \\
	& \quad + (4-4\mu^2+3\mu^4)F(\mu)\big]\\
\gamma_4(\mu) & = \frac{9}{4} F(\mu)\\
\end{split}
\label{gammamuhpar.eq}
\end{equation}

$\bullet$ In the $\hperp$ channel:
\begin{equation}
\begin{split}
\gamma_1(\mu) & = \frac{3}{256(1-\mu^2)^4}
	\big[ -2(80-56\mu^2+42\mu^4+39\mu^6) \\
	& \quad +(122-144\mu^2+127\mu^4)\sqrt{1+\mu^2} \\
	& \quad + 3(32-64\mu^2+96\mu^4-48\mu^6+19\mu^8)F(\mu)\big]
\end{split}
\label{gammamuhperp.eq}
\end{equation}

$\bullet$ In the $\lpar$ channel:
\begin{equation}
\begin{split} 
\gamma_1(\mu) & = \frac{3}{512(1-\mu^2)^4}
	\big[ -288+48\mu^2-252\mu^4-138\mu^6 \\
	& \quad + (222-144\mu^2+237\mu^4)\sqrt{1+\mu^2}\\
	& \quad + (192-384\mu^2+720\mu^4-336\mu^6+123\mu^8)F(\mu)\\
	& \quad +  A_1(\mu) \cos 2\phi + A_2(\mu)  \cos 4\phi \big] \\
\gamma_3(\mu) & = \frac{9}{16(1-\mu^2)^2}
	\big[ -4-2\mu^2 + 3\sqrt{1+\mu^2} \\
	& \quad + (4-4\mu^2+3\mu^4)F(\mu) 
                + B(\mu) \cos 2\phi \big] \\
\gamma_4(\mu) & = \frac{9}{4} F(\mu)
\end{split}
\label{gammamulpar.eq}
\end{equation}

$\bullet$ In the $\lperp$ channel:
\begin{equation}
\begin{split}
\gamma_1(\mu) & = \frac{3}{512(1-\mu^2)^4}
	\big[ 32-176\mu^2-84\mu^4+18\mu^6 \\
	& \quad + (-22+144\mu^2-17\mu^4)\sqrt{1+\mu^2}\\
	& \quad + 3\mu^4(48-16\mu^2+3\mu^4)F(\mu) \\
	& \quad - A_2(\mu) \cos 4\phi  \big] \\
\gamma_2(\mu) & = \frac{9}{8(1-\mu^2)^2}
	\big[ -4-2\mu^2 + 3\sqrt{1+\mu^2}\\
	& \quad + (2+\mu^4)F(\mu)\big]
\end{split}
\label{gammamulperp.eq}
\end{equation}
All other $\gamma_i$ are zero as evident from (\ref{cofxtable}). 
In the linear channels, the anisotropic contributions from 
(\ref{anisotropies.eq}) are 
weighted by  
\begin{equation}
\begin{split}
A_1(\mu) & = \big[ -56(-2+8\mu^2+4\mu^6+5\mu^8) \\
 	& \quad +28 (-4+18\mu^2-14\mu^4+15\mu^6)\sqrt{1+\mu^2} \\
 	& \quad + 12\mu^4(16+8\mu^2+6\mu^4+5\mu^6)F(\mu)\big]/\mu^2\\
A_2(\mu) & = \big[ 48-152\mu^2+128\mu^4+48\mu^6-212\mu^8-70\mu^{10}\\
	&+ (-48+176\mu^2-222\mu^4+88\mu^6+111\mu^8)\sqrt{1+\mu^2} \\
	& \quad +3\mu^8(8+24\mu^2+3\mu^4)F(\mu) \big] /\mu^4\\
B(\mu)& = [ 2-4\mu^2-4\mu^4+(-2+5\mu^2)\sqrt{1+\mu^2}\\
	& \quad +\mu^4(2+\mu^2)F(\mu) ]/\mu^2
\end{split}
\end{equation}
In all these expressions, the auxiliary function $F(\mu)$ is given by:
\begin{equation}
F(\mu) =2 \arg\cosh \left(\frac{1}{|\mu|}\right)- 
\arg\cosh \left(\frac{1}{\mu^2}\right) 
\end{equation}
Under this form, $F(\mu)$ is not a manifestly real function
of $\mu.$ It can be rewritten as:
\begin{equation}
\begin{cases}
\displaystyle \frac{2}{\sqrt{1-\mu^2}} \arg\sinh\left( \frac{\sqrt{1+\mu^2}-1}{\sqrt{2}\mu^2}
 \sqrt{1-\mu^2} \right) &, |\mu|<1 \\ 
\displaystyle \frac{2}{\sqrt{\mu^2-1}} \arcsin \left( \frac{\sqrt{1+\mu^2}-1}{\sqrt{2}\mu^2}
 \sqrt{\mu^2-1} \right) &, |\mu|>1 
\end{cases}
\end{equation}
In table (\ref{cofxtable}), the function $c_4(x)$ is just a constant,
without any angular dependence on $\phi$ or $x=\cos \vartheta$. The
corresponding contribution $\gamma_4(\mu)$ in (\ref{gammamuhpar.eq}) and
(\ref{gammamulpar.eq}) is essentially $F(\mu)$.  
We see therefore that   
$F(\mu)$ is -- within a factor 9/4 -- 
the crossed double-scattering bistatic coefficient
for a scalar wave 
scattered by a semi-infinite homogeneous medium of
point scatterers. 
It is a bell-shaped function around $\mu=0$
with width of the order of unity.
  
With the help of the previous expressions, the scattered intensity can
be plotted, for all atomic transitions, all polarization channels and
all directions of observation.

\subsection{Crossed contribution for exact backscattering}

In exactly the backscattering direction $\mu=0$, the
above expressions simplify
considerably, yielding the crossed bistatic coefficient in
the backscattering direction
\begin{equation}
\begin{split}
\gamma_{\mathrm C2}(0) & = \frac{9}{8}\big(c_1(w_1+w_3)^2 + c_2 w_1w_3
\\ &\quad 
+ c_3(w_1+w_3)w_2 + c_4 w_2^2 \big) 
\end{split}
\end{equation}    
where the numerical coefficients $c_i$ are given as functions of the
polarization channels:
\begin{center}
  \renewcommand{\arraystretch}{1.3}
\begin{tabular}{|c|c|c|c|c|}
\hline
    &$\hpar$            &  $\hperp$     &$\lpar$        &$\lperp$       \\
\hline
$c_1$ &$\frac{5}{48} $&$\ln 2-\frac{19}{48}$ &
    $\ln 2-\frac{11}{32}$ & $\frac{5}{96}$\\
$c_2$ &$2\ln 2-1$               &$0$            &       $0$&$2\ln 2-1$  \\
$c_3$ &$2\ln 2-\frac{1}{2}$&$0$ &$ 2\ln 2-\frac{1}{2}$&$0$      \\
$c_4$ &$2\ln 2$         &$0$    &$2\ln 2$ & $0$ \\
\hline
\end{tabular}
\renewcommand{\arraystretch}{1.0}
\end{center}
Just as for the ladder contribution, the coefficient $c_1$ had been
derived in ref.~\cite{vanTiggelen90}, the others describe the
generalization to the case of degenerate atomic transitions.  In the
parallel channels, all crossed coefficients $c_i$ are equal to the  
corresponding ladder
coefficients $l_i$ (\ref{laddercoeff12.eq}-\ref{laddercoeff34.eq}).
This is the signature of reciprocity since
the ladder and crossed contributions are then equal for $w_2=w_3$. In
the perpendicular channels, no such correspondence can be observed.

\subsection{Wings of the crossed contribution}
\label{wings.sec}

For a large reduced scattering angle $\mu=k\ell\theta \gg 1$, the
previous expressions can be expanded in powers of $\mu^{-1},$
giving the wings of the enhanced backscattering peak. 
This asymptotic expression describes the wings of the backscattering
peak even if higher orders of scattering contribute to the intensity
at smaller angles. 
The crossed bistatic coefficient in the wings becomes
\begin{equation}
\begin{split}
  \gamma_{\mathrm C}(\mu) & = \frac{9\pi}{8\mu}
  \big( a_1(w_1+w_3)^2 + a_2 w_1w_3 
\\ & \quad  
  + a_3(w_1+w_3)w_2 + a_4 w_2^2 \big) + {\mathcal O}(\mu^{-2})
\end{split}
\end{equation}
where the wing coefficients are
\begin{center}
  \renewcommand{\arraystretch}{1.3}
\begin{tabular}{|c|c|c|c|c|}
\hline
    &$\hpar$            &  $\hperp$     &$\lpar$        &$\lperp$       \\
\hline
$a_1$ &$\frac{3}{64} $&$\frac{19}{64}$ &
  $\frac{1}{16}(3+2\cos^2\phi+3\cos^4\phi)$ & $\frac{3}{64}\sin^2 2\phi$\\
$a_2$ &$\frac{1}{2}$            &$0$    &$0$            & $\frac{1}{2}$\\
$a_3$ &$\frac{3}{4}$&$0$        &$\frac{1}{2}(1+\cos^2\phi)$&$0$        \\
$a_4$ &$1$              &$0$    &$1$ & $0$      \\
\hline
\end{tabular}
\renewcommand{\arraystretch}{1.0}
\end{center}
The wing coefficients in the linear channels depend
on the angle $\phi$ between the incident polarization and the
direction of the intensity scan, carrying the anisotropy of the linear
backscattering peaks. 
In the $\lpar$ channel, the intensity is higher in the direction of
the polarization ($\phi=0$) than perpendicular to it, yielding a 
cigar-shaped intensity pattern in the plane of observation (in
eq.~(A.5) of ref.~\cite{vanTiggelen90}, 
a term $-\frac{3}{4}\sin^22\phi$ is missing, 
otherwise all coefficients $a_1$ coincide).  
In the $\lperp$ channel, the intensity is smaller in the directions of
incident ($\phi=0$) and scattered ($\phi=\pi/2$) polarization than
along the diagonals, yielding a cloverleaf pattern in the observation
plane. 
As pointed out in ref.~\cite{vanTiggelen90}, in
the $\lperp$ channel the classical coefficient $a_1 \propto \sin^2
2\phi$ vanishes if the intensity is scanned in the direction parallel
or perpendicular to the incident polarization
($\phi=0,\frac{\pi}{2}$).  That means that the peak decreases in these
directions as $\mu^{-2}$ instead of $\mu^{-1}$. 
But for atoms, a
second constant coefficient $a_2=\frac{1}{2}$ comes into play that
maintains a (modulated) decrease as $\mu^{-1}$ in all directions, thus
reducing the anisotropy
In the $\lpar$ channel, the above expressions
permit to verify that the classical anisotropy ratio
$\gamma_{\mathrm C}(\phi=0)/\gamma_{\mathrm C}(\phi=\frac{\pi}{2})=\frac{8}{3}$ decreases
as the atomic degeneracy increases, converging to $\frac{40}{19}$ for
transitions of type $\Je=J\pm 1$ and to $\frac{40}{22}$ for transitions
of type $\Je=J$ as $J\rightarrow\infty$.

\end{document}